\documentclass[twocolumn]{aastex631} 

\usepackage{amssymb,amsmath,graphicx,natbib,hyperref}
\usepackage{multirow}
\usepackage{subfigure}
\usepackage{float}

\renewcommand{\thefootnote}{\fnsymbol{footnote}}
\usepackage{color}
\usepackage[flushleft]{threeparttable}
\usepackage{enumitem}
\usepackage{natbib}
\usepackage{tabularx}
\usepackage{booktabs}

\newcommand{\asec}{$^{\prime\prime}$}
\newcommand{\kms}{\,km\,s$^{-1}$}

\newcommand{\um}{\,$\mu$m}

\newcommand{\el}		  {\mbox{$E_l$}}
\newcommand{\gl}		  {\mbox{$g_l$}}

\newcommand{\nl}		  {\mbox{$N_l$}}
\newcommand{\aij}		  {\mbox{$A_{ij}$}}
\newcommand{\mo}          {\mbox{$^{-1}$}}
\newcommand{\vLSR}        {$v_{LSR}$}
\newcommand{\tauu}		  {$\tau$}
\newcommand{\nuu}		  {$\nu$}

\newcommand{\cmms}		  {cm$^{-2}$}
\newcommand{\s}		  {$\sim$}

\newcommand \percmsq {\,cm$^{-2}$}
\newcommand \percmcu {\,cm$^{-3}$}

\newcommand \co {$^{12}$CO}
\newcommand \thco {$^{13}$CO}
\newcommand \ceio {C$^{18}$O}
\newcommand{\cso}         {\mbox{C$^{17}$O}}

\newcommand{\tex}		  {$T_\mathrm{ex}$}
\newcommand{\tk}		  {$T_\mathrm{k}$}

\newcommand{\wt}        {W3~IRS5}

\shortauthors{Li et al.}

\shorttitle{M-band observations of W3~IRS5 with \textit{iSHELL/IRTF} }

\begin{document}

\title{High-Resolution M-band Spectroscopy of  CO towards the Massive Young Stellar Binary \mbox{W3 IRS5}}

\correspondingauthor{Jialu Li}
\email{jialu@astro.umd.edu}

\author[0000-0003-0665-6505]{Jialu Li}
\affiliation{Department of Astronomy, University of Maryland, College Park, MD 20742, USA}

\author[0000-0001-9344-0096]{Adwin Boogert} 
\affiliation{Institute for Astronomy, University of Hawaii, 2680 Woodlawn Drive, Honolulu, HI 96822, USA}

\author[0000-0003-4909-2770]{Andrew G. Barr}
\affiliation{Leiden University, Niels Bohrweg 2, 2333 CA Leiden, The Netherlands}

\author{Alexander G. G. M. Tielens}
\affiliation{Department of Astronomy, University of Maryland, College Park, MD 20742, USA}
\affiliation{Leiden University, Niels Bohrweg 2, 2333 CA Leiden, The Netherlands}

\received{}
\revised{}
\accepted{}
\submitjournal{ApJ}

\begin{abstract}

We present in this paper the results of high spectral resolution ($R$=88,100) spectroscopy at 4.7~\um\ with iSHELL/IRTF of hot molecular gas close to the massive binary protostar \wt. The binary was spatially resolved and the spectra of the two sources (MIR1 and MIR2) were obtained simultaneously for the first time. Hundreds of \co\ \nuu=0--1, \nuu=1--2 lines, and \nuu=0--1 transitions of the isotopes of \co\ were detected in absorption, and are blue-shifted compared to the cloud velocity $v_{LSR}=-$38~\kms. We decompose and identify kinematic components from the velocity profiles, and apply rotation diagram and curve of growth analyses to determine their physical properties. Temperatures and column densities of the identified components range from 30--700~K and 10$^{21}$--$10^{22}$~\percmsq, respectively. {Our curve of growth analyses consider two scenarios. One assumes a foreground slab with a partial covering factor, which well reproduces the absorption of most of the components. The other assumes a circumstellar disk with an outward decreasing temperature in the vertical direction, and reproduces the absorption of all the hot components.} We attribute the physical origins of the identified components to the foreground envelope ($<$100~K), post-J-shock regions (200--300~K), and clumpy structures on the circumstellar disks ($\sim$600~K). We propose that the components with a J-shock origin are akin to water maser spots in the same region, and are complementing the physical information of water masers along the direction of their movements.

\end{abstract}
    
\section{Introduction}

Although massive stars profoundly affect the evolution of the Universe, their formation and evolution processes are not well-understood. Massive stars are rare, deeply embedded in the early stage, and are seldom found to form in isolation. Therefore the large distances to the observers, the high extinction at optical and near-infrared wavelengths, and the highly clustered environment impede a clear understanding of their formation and evolution processes. 

Theoretical models for massive star formation have remained controversial. {Compared to the well-established formation process of low-mass stars \citep{mc07}, massive stars do not form through an exact scaled-up mechanism due to the strong radiation pressure, which dramatically influences the accretion rate and the final stellar mass \citep{wolfire87}.
Several approaches have been followed to overcome this problem: the generation of radiatively driven bubbles and the disc-mediated accretion \citep{krum09, rosen20} in monolithic collapse models \citep{mt03, krum05} have been developed as a way to overcome the radiation pressure barrier; the coalescence scenario \citep{bonnell98, bally05} in high stellar density environments avoids the radiation pressure issues; the competitive accretion model \citep{bonnell04, bonnell06} suggests that the forming stars accrete material that is not gravitationally bound to the stellar seed. Each of these different scenarios has implications for cluster formation and binary formation
involving {disks}.} 

For high-mass star-forming cores, the current proposed theoretical evolutionary sequence is: high-mass starless cores (HMSCs) $\rightarrow$ high-mass cores harboring accreting low/intermediate-mass protostar(s) destined to become a high-mass star(s) $\rightarrow$ high-mass protostellar objects (HMPOs) $\rightarrow$ final stars \citep{beither07}. Observationally, the embedded phases of massive protostellar objects are subdivided into infrared dark clouds (IRDC), hot molecular cores (HMCs), hypercompact- and ultracompact-HII regions (HCHIIs and UCHIIs), and compact and classical HII regions \citep{beither07}. As the formation and evolution proceed, the central object warms and ionizes the environment, and drives a rich chemistry. Complex physical activities are involved in the evolution as well, such as accretion disks, outflows, shocks, and disk winds \citep{cesaroni07, zin07}. 

In the proposed evolutionary sequence of massive star formation, each stage has its own characteristic physical conditions. {Mid-infrared (MIR) spectroscopy is sensitive to the presence of warm gas (several hundreds of degrees) that is very close to the protostar, often at a distance between 100--1000 AU.} Observing at mid-IR wavelengths, therefore, fills the gap in between the cooler and more extended regions ($>\mbox{1000 \textrm{AU}}$) emitting in the submm/millimeter and the innermost ionized HII regions traced by observations at radio wavelengths. Mid-IR spectroscopy also traces important characteristic chemistry during massive star formation. At these high temperatures, grain mantles will have sublimated and neutral-neutral reaction channels have opened up, resulting in a rich inventory of organic molecules \citep{vdt03, agundez08, herbst09, bast13}.

Molecular ro-vibrational transitions in the mid-IR provide a unique opportunity to study the physical conditions and the chemical inventory of embedded phases in massive star formation. {The size of the mid-IR continuum emission region provides the effective spatial resolution of such spectroscopic observations because the observed absorption components are exactly located in front of the infrared source and are along the line of sight.} The full set of ro-vibrational lines can be covered in a short bandwidth without multiple frequency settings that sub-millimeter observations require. {Molecules without dipole moments such as C$_2$H$_2$ and CH$_4$, which are among the most abundant carbon-bearing molecules, can only be observed through their ro-vibrational spectra in infrared.} Therefore, mid-IR spectroscopy at high resolution allows us to study the properties of physical components close to massive proto-stars, and to understand the interactions of the massive protostars with their environment in a better way.

\wt\ is an active star-forming region in the Perseus arm at a distance of {2.3$^{+0.19}_{-0.16}$~kpc \citep[Gaia-DR2;][]{navarete19}}. 
The high IR luminosity and the presence of radio sources reveal the presence of high-mass protostars.  \wt\ is a binary {\citep{megeath96}} and we refer to the northeastern component as MIR1 and the southwestern one as MIR2, following the nomenclature in \citet{vdt05}.  Near-IR images reveal that MIR1 and MIR2 are separated by $\sim$1.2\asec\ and are coincident with the bright sub-mm sources, MM1 and MM2 \citep{vdt05, megeath05}. In this paper, we present a rich high-resolution spectrum of W3 IRS5 in the 4.7~\um\ $M$-band, covering ro-vibrational transitions of \co\ and its isotopologues \thco, \ceio, \cso. In contrast to early observations by \citet{mitchell91} at the same wavelength, MIR1 and MIR2 are now spatially resolved, and we are therefore able to separate the different kinematic components in the complex absorption line profiles, tracing the immediate environment of each source in the \wt\ binary. We describe {our} observations and data reduction in Section~\ref{sec:obs}, and our analysis method includes a simple optically thin foreground as well as a photospheric disk model slab model in Section~\ref{sec:methods}. We present the identification process and the derived physical conditions of different kinematic components in Section~\ref{sec:results}, and discuss the implications of our observations for our understanding of high-mass star formation in W3 IRS 5 in Section~\ref{section: discussion}.

\section{Observations and data reduction} \label{sec:obs}

\begin{figure*}[!t]
    \centering

    {\includegraphics[width=\textwidth]{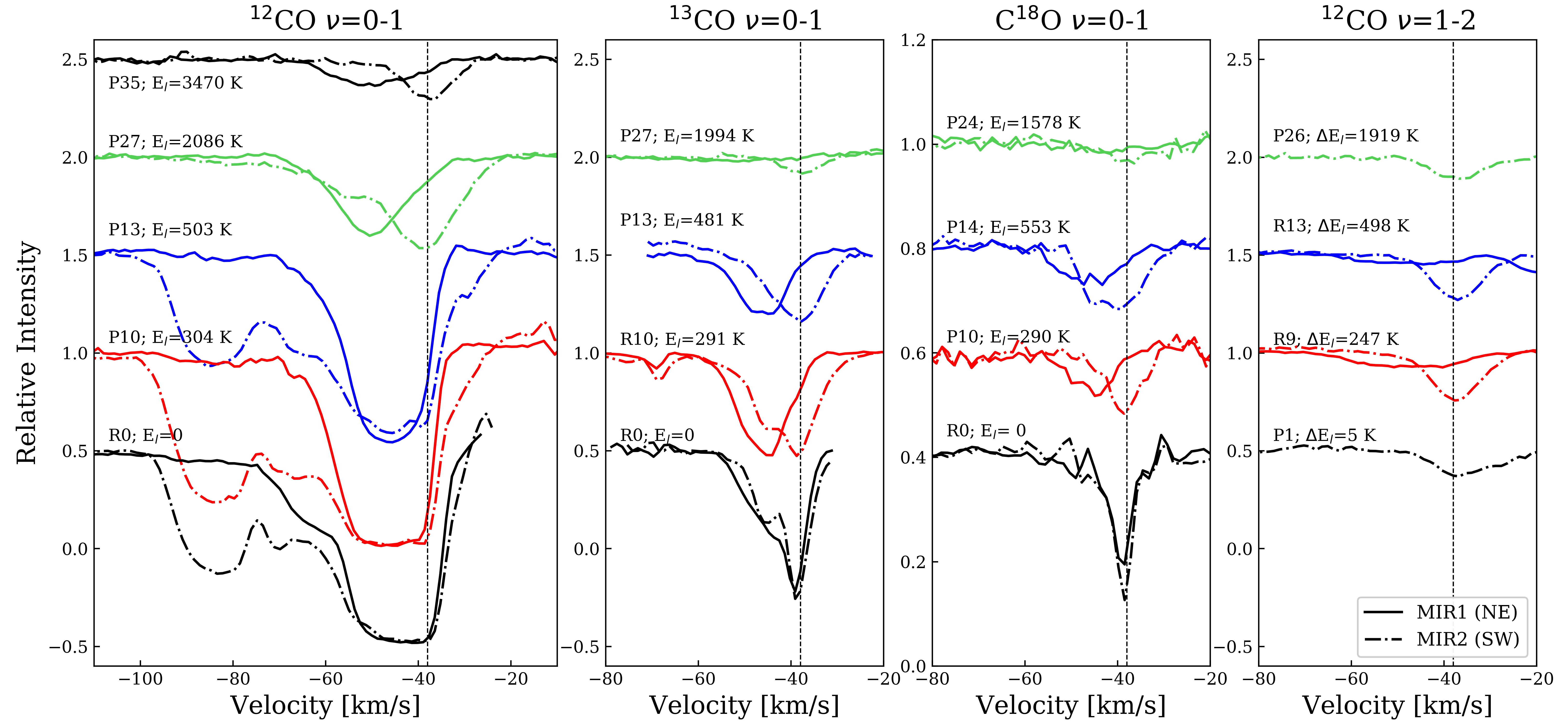}  }

    {\caption{Selected \co, \thco, \ceio\ \nuu=~0--1, and \co\ \nuu=~1--2 absorption lines observed towards MIR1 (\textit{solid}) and MIR2 (\textit{dash-dotted}). The dashed vertical lines at -38~\kms\ are the systematic velocities. In the panel of \co\ \nuu=~1--2, $\Delta E_l$ = $E_{\nu=1, J=J_l}-E_{\nu=0, J=0}$. Transitions with similar energy levels are represented by the same color. \co\ \nuu=~1--2 P1 and P26 of MIR1 are not plotted due to their poor spectral quality. \cso\ \nuu=0--1 spectra were not plotted because of the limited energy levels of the observed lines. We note that the absorption at -60~\kms\ on \co\ \nuu=0--1 R0 of MIR1 is contaminated by \thco\ \nuu=0--1 R14, and we do not use the two transitions in our analysis.}\label{fig:com1}}
    
\end{figure*}

We observed W3 IRS5 with the iSHELL spectrograph \citep{rayner22} at the NASA InfraRed Telescope Facility (IRTF) 3.2-meter telescope as part of program 2018B095 on UT 09:00 2018 October 5. The instrument was used in its {spectral mode M1 \citep[see Table 1 in][]{rayner22}} with a slit width of 0.375\asec. This provides a resolving power of $R=\lambda/\Delta\lambda=88,100\pm2,000$ \citep{rayner22} over a wavelength range of 4.52--5.25 \um, excluding small gaps between the echelle orders. The total on-source integration time was 30 minutes, and the airmass was in the range of 1.535--1.443. The 15\asec\ long slit was oriented along a position angle of 37 degrees, so that the binary components of \wt\ were observed simultaneously. The seeing conditions allowed for the 1.2\asec\ binary to be well separated in the M-band. The targets were nodded along the slit,  allowing for the subtraction of the sky and hardware background emission. The Spextool package \citep[version 5.0.2,][]{cushing04} was used to reduce the spectra. This includes wavelength calibration using the sky emission lines, and custom extraction apertures to separate the binary components. The binary components are of similar brightness in the M-band. {In the extracted spectra, the contamination by the flux from the other binary component is no more than $\sim$~5--7$\%$. This is estimated from the spectral features at velocities $v_{\rm LSR}<-70$ km/s (Figure~\ref{fig:com1}), where we assume MIR1 only has continuum and the absorption lines occur exclusively in MIR2.} Telluric absorption lines were divided out using the program Xtellcor\_model\footnote{\url{http://irtfweb.ifa.hawaii.edu/research/dr\_resources/}}, which makes use of atmospheric models  calculated by the Planetary Spectrum Generator \citep{vill18}. {The echelle orders of iSHELL are strongly curved (blaze shape; cf., Figure~8 in iSHELL's observing manual\footnote[6]{\url{http://irtfweb.ifa.hawaii.edu/\~ishell/iSHELL\_observing\_manual\_20210827.pdf}}). This was corrected for by dividing by flat field images taken with iSHELL’s internal lamp.} The Doppler shift due to the combined motion of the Earth on the date of the observations and the systemic velocity of W3 IRS5 \citep[$v_{\rm LSR}$ =$-38$~\kms;][]{vdt00} is $-56$~\kms. This is sufficient to separate the deep telluric CO lines from those in \wt. Residual baseline curvature was divided out using a median filter. We shifted the wavelength scale by $-18$~\kms\ to remove the  motion of the Earth in the direction of \wt, converting it to an LSR scale. Finally, we used the HITRAN database \citep{kochanov16} to identify the rovibrational transitions of \co\ and its isotopologues. 

W3 IRS5 was also
observed with the SpeX spectrometer \citep{rayner03} at the IRTF in order to obtain a wider
wavelength view of this binary system. The observations were done on UT 14:00 2020 August
14. The 15\asec\ long SpeX slit was oriented along the binary position angle of 37 degrees,
and guiding was done in the $K$-band on the slit spill-over flux. Spectra were taken with the
SpeX LXD\_Long mode, using the 0.5\asec\ wide slit. This yields a resolving power of
$R=1,500$. The instantaneous spectral coverage is 1.95--5.36\um. The standard star was HR
1641 (B3V). The IRTF/SpeX spectra were reduced using Spextool version 4.1 \citep{cushing04}. Flat fielding was done using the images obtained with SpeX's calibration
unit. The wavelength calibration procedure uses lamp lines at the shortest wavelengths and sky
emission lines in much of the $L$ and $M$-bands. At the good seeing of $\sim$0.5\asec, the
binary was well separated, and could be extracted without significant contamination. The
telluric correction was done using the Xtellcor program \citep{vacca03}. This uses a model
of Vega to divide out the stellar photosphere. Vega's spectral type A0V differs from that of the
standard star, and thus care must be taken with interpreting features near hydrogen lines.

\section{Methods}\label{sec:methods}

For both MIR1 and MIR2, we have detected several hundreds of lines in the \nuu=0--1 band of \co, \thco, \ceio, \cso, and the \nuu=1--2 band of \co \footnote[1]{In the rest of the paper we consider \nuu=0--1 as the default band of \co\ unless `\nuu=1--2' is specified.} at 4.7~\um\ in absorption. Figure~\ref{fig:com1} shows a selected group of lines that illustrate that each MIR source has a number of distinctive kinematic components that are characterized by different excitation conditions. For example, MIR2 shows highly blue-shifted gas, up to -90 \kms, but MIR1 does not. The \thco, \cso, \ceio\ lines of MIR2 are centered on the systemic velocity of $v_{\rm LSR}=-38$~\kms. In contrast, for MIR1 lines of the lowest J-levels center on $v_{\rm LSR}=-38$~\kms, while the high-J lines center on $-46$~\kms. To explore the properties of these components, we analyze the rovibrational lines of each species simultaneously. We regard the optically thin slab model in LTE as an appropriate start for optically thin analysis (Section~\ref{subsec:lte}). In Section~\ref{subsec:ana}, we apply corrections when effects of optical depth, covering factor, and radiative transfer are important.

\subsection{Preliminary Analysis: Optically Thin Slab Model under LTE}\label{subsec:lte}

Recovering the column density information from an absorption line {is} straightforward, if emission from the molecular gas is negligible, and the relative intensity of the line to the continuum can be described by an attenuation factor, e$^{-\tau}$, in which $\tau$ is defined as the optical depth. This is the commonly considered slab model, where a cold, isothermal absorbing cloud is in front of a hot continuum source. In the context of the direct environment of a massive protostar, the foreground cloud can absorb the mid-IR emission from a disk or Hot Core. In the context of this paper, we note that the Planck function peaks at 4.7~$\mu$m for a temperature of ~600 K. Hence, the background continuum source will have to have a temperature of that order or higher. Moreover, in order to see absorption lines, the continuum source has to have an emission optical depth at least of order 1, while the foreground cloud has to be considerably cooler than 500 K (Barr et al 2022, submitted). 

In a slab model, when the lines are optically thin, we can get the column density $N_l$ in the lower state of a transition directly from the integrated line profile by

\begin{equation}
N_l = 8\pi/(A_{ul} \lambda^3) g_l/g_u \int \tau(v)dv, \label{eq:1}
 \end{equation}
in which $A_{ul}$ is the Einstein coefficient, $g_l$ and $g_u$ are the statistical weight of the lower and upper level, and
\begin{equation}
\tau(v) = -\textrm{ln}(I_\textrm{obs}/I_c),\label{eq:2}
\end{equation}
where $I_\textrm{obs}$ and $I_c$ are the intensity of the absorption and the continuum. 

If the absorbing gas is in LTE at an excitation temperature, \tex, the population in the rotational level $J$ is thermalized according to 
\begin{equation}
    \frac{N_l}{g(J)} = \frac{N_\textrm{tot}}{Q(T_\textrm{ex})} \textrm{exp}\left(-\frac{E_l}{k_B T_\textrm{ex}}\right), \label{eq:bd}
\end{equation}
where $N_\textrm{tot}$ is the total column density, $E_l$ is the excitation energy, $g(J)$ is the statistical weight of the level ($g(J)=2J+1$ for a linear molecule), and $Q(T)$ is the partition function. For a uniform excitation temperature, the rotation diagram, ln($N_l/(2J+1)$) versus $E_l/k_B$, follows a straight line, with the inverse of the slope representing the temperature, and the intercept representing the total column density over the partition function. We can therefore derive the temperature and the total column density of the molecular gas. If the slope on the rotation diagram is not a constant but has a gradient, we regard the component as a compound of multiple temperatures, and fit ln($N_l/(2J+1)$) versus $E_l/k_B$ with 
\begin{equation}
    \frac{N_l}{2J+1} = \sum_i \frac{N_{\textrm{tot,}\,i}}{Q(T_{\textrm{ex,}\,i})} \textrm{exp}\left(-\frac{E_l}{k_B T_{\textrm{ex,}\,i}}\right), \label{eq:bd2}    
\end{equation}
where `\textit{i}' represents the `\textit{i}-th' temperature component. 

\subsection{Curve of Growth Analysis} \label{subsec:ana}

\subsubsection{Slab Model of a Foreground Cloud}

For an absorbing foreground slab, corrections for the line saturation are necessary for optically thick lines. We can use the measured equivalent width,
\begin{equation}
W_\lambda = \int (1 - I_\textrm{obs}/I_c) d\lambda = \int (1 - \textrm{e}^{-\tau}) d\lambda,
\end{equation}
to obtain the column density of each state from the curve of growth \citep{rodgers74}:

\begin{equation}
    \frac{W_\lambda}{\lambda} \approx 
        \begin{cases}
      \frac{\sqrt{\pi}b}{c}\frac{\tau_p}{1+\tau_p/(2\sqrt{2})}&  \text{for} ~\tau < 1.254\\
      \frac{2b}{c}\sqrt{\textrm{ln}(\tau_p/\textrm{ln}2) + \frac{\gamma \lambda}{4b\sqrt{\pi}}(\tau_p - 1.254)} &  \text{for} ~\tau > 1.254
    \end{cases} , \label{eq: draine}
\end{equation}
where the peak optical depth is given by 

\begin{equation}
    \tau_p = \frac{\sqrt{\pi}e^2}{m_e bc}N_{l}f_{lu}\lambda. \label{drain-tau}
\end{equation}
In the equations above, $f_{lu}$ is the oscillator strength, and $\gamma$ is the damping constant of the Lorentzian profile. For CO rovibrational lines, $\gamma$ due to radiative damping is of order $\sim$10 s$^{-1}$. The Doppler parameter in velocity space, $b$, is related to the full width at half maximum of an optically thin line by $\Delta v_{\textrm{FWHM}} = 2\sqrt{\textrm{ln}2}b$.  We stress that the Lorentzian line width that $\gamma$ corresponds to ($10^{-9}$~\kms) is negligible compared to the observed Doppler width (a few~\kms). 

As observations revealed that strong absorption lines did not go to zero intensity, \citet{lacy13} recognized several issues that require cautions when applying an absorbing slab model. One is that if emission from the foreground molecular cloud is not negligible, the line intensity tends to approach the source function, and does approach the source function at a sufficient optical depth. The source function equals the Planck function at the line wavelength if the molecular gas is at LTE, which requires sufficient density if no other scattering opacity is considered inside the molecular cloud. As a reference, for a representative background temperature above 600~K, the foreground emission contributes a $\sim$ 4$\%$ residual intensity at \mbox{$\tau\sim$~5} in a 400~K cloud. The emission is therefore negligible in cooler clouds with smaller columns. 

Another problem
that may occur is that the foreground cloud does not cover the entire observing beam. The absorption feature saturates at a non-zero intensity as well because of the dilution, even if the emission from the gas is not important. Should a covering factor $f_c$ be considered, equation (\ref{eq:2}) is modified to
\begin{equation}
    I_{\textrm{obs}} = I_{\textrm{c}} (1 - f_c (1 - \textrm{e}^{-\tau(v)})). \label{eq:ff}
\end{equation}
Similarly,  the left-hand side of equation \ref{eq: draine} is modified to $W_\lambda/(\lambda f_c)$.

\subsubsection{Stellar Atmosphere Model of a Circumstellar Disk}\label{mode:sam}

The absorption may also occur if the dust thermal continuum is mixed with the molecular gas, and there is an outward-decreasing temperature gradient. This scenario is similar to the stellar atmosphere model when the continuum and the line are coupled, in which \textcolor{black}{the residual flux}, 
\begin{equation}
    R_\nu \equiv I_\nu/I_c,
\end{equation}
can then be approximated by the Milne-Eddington model \citep[][Ch 10]{mihalas78} which assumes a grey atmosphere. In the system of a forming massive star, such a model can be realized in a circumstellar disk that has a heating source in the mid-plane. In this scenario, saturated absorption lines approach a constant depth and there is no need to consider a covering factor.  We refer to Appendix A in \citet{barr20} for details of the expected line residual flux in this model.

Following \citet{mihalas78}, for the stellar atmosphere model, when there is pure absorption in the lines, the curve of growth is constructed by considering the equivalent width versus $\beta_0$, the ratio of the line opacity at line center, $\kappa_L(\nu=\nu_0)$, to the continuum opacity $\kappa_c$:
 \begin{equation}\label{eq: mihalas}
  \begin{aligned}
\frac{W_\nu}{2\Delta \nu_D} &= \int_0^{+\infty} (1-R_v) dv \\
     &= A_0\int_0^{+\infty} \beta_0 H(a, v)[1+\beta_0 H(a, v)]^{-1}dv, 
 \end{aligned}
\end{equation}
in which
\begin{equation}
 \frac{W_\nu}{2\Delta \nu_D} = \frac{W_v}{2b} = \frac{c}{2b}\frac{W_\lambda}{\lambda},
\end{equation}
and 
 \begin{equation}\label{eq:beta0}
  \begin{aligned}
   \beta_0 &= \frac{\kappa_L(\nu=\nu_0)}{\kappa_c}\\ &= \frac{A_{ul}\lambda^3}{8\pi\sqrt{2\pi}\sigma_v}\frac{g_u}{g_l}\frac{N_l}{\sigma_c N_H} \left(1-\frac{g_l}{g_u}\frac{N_u}{N_l}\right).
 \end{aligned}
\end{equation}

In the equations above, $v$ is the frequency shift with respect to the line center in units of the Doppler width, $H(a, v)$ is the Voigt function that gives the line profile, in which the damping factor $a=\gamma \lambda/b$ is of the order of 10$^{-8}$ for CO ro-vibrational lines. The parameter $A_0$ is the central depth of an opaque line, and its exact value is determined by the radiative transfer model of the surface of the disk. The value of $A_0$ is related to the gradient of the Planck function, $dB/d\tau_c/B(T_o)$, where $B$ is the Planck function, $\tau_c$ is the continuum optical depth, $T_o$ is the surface temperature of the disk. For a grey atmosphere, $A_0$ is \mbox{$\sim$ 0.5--0.9} from 900 to 100 K \citep[see Appendix A in][]{barr20}. The dispersion in velocity space, \textcolor{black}{$\sigma_v$}, is transformed from the Doppler parameter, $b/\sqrt{2}$. The continuum opacity, $\kappa_c$, is given by the dust cross-section per H-atom $\sigma_c$. We adopt a value of \mbox{7$\times 10^{-23}$~cm$^2$/H-nucleus} for $\sigma_c$ following \citet{barr20}, as it is appropriate for coagulated interstellar dust \citep{ormel11}. We can eliminate the bracketed item in \mbox{equation \ref{eq:beta0}} if stimulated emission is negligible.

Similarly, for molecular gas under LTE, we may express $\beta_0$ as

\begin{equation}
        \beta_0 = \left( \frac{A_{ul}\lambda^3}{8\pi\sqrt{2\pi}\sigma_v \sigma_c}\frac{g_u}{g_l} \right)\frac{g_l\textcolor{black}{N_{tot}}}{Q(\textcolor{black}{T})\textcolor{black}{N_H}}e^{-E_l/(k\textcolor{black}{T})},
    \end{equation}
where $Q(T)$ is the partition function, and $N_{tot}/N_H$ is the relative abundance of the molecules to hydrogen. If LTE sustains, we can thereby retrieve $(N_{tot}/N_H, T)$ through a grid search method by comparing the observable ${W_\nu}/{2\Delta \nu_D}$ (or transform into ${W_\lambda}/{\lambda}$) with the theoretical curve of growth and looking for the smallest $\chi^2_r$. We note that our choice of $\sigma_c$ influences the derived absolute abundance, although we may still use the derived abundance to calculate the relative abundance of different species in the same kinematic component.

\subsubsection{Comparing the Two Curve of Growth Analyses}

Although the two curve of growth analyses assume intrinsically different radiative transfer models, the absorption profiles evolve in a similar way as the optical depth increases. The line profile firstly grows like a Gaussian (the linear part), and then saturates the intensity at the line center and thus increases the equivalent width slowly through absorption in the (Gaussian) wings (the logarithmic part). Finally, the equivalent width grows quickly again when the Lorentzian wing takes over (the square root part). The latter case does not apply to the physical conditions in this paper because the Lorentzian parameters $\gamma$ and $a$ (see \S~\ref{mode:sam}) are too small.

The main difference between the two models exists in the lower limit of the center depth of a line. In the stellar atmosphere model, $A_0$ does not approach zero; in a slab model with a 100$\%$ covering factor, the line depth does saturate at zero intensity. This is due to a mixture of the origin of the absorption line and the continuum, and results in a difference in the equivalent width. However, for a slab model with a partial covering factor, its curve of growth may be alike to that of the stellar atmosphere model under certain conditions.

To illustrate this point, we first examine both models in the optically thin limit. In the foreground slab model, $W_\lambda/\lambda$ goes to $f_c \tau_p$ (see eqn~\ref{eq: draine}), and in the atmosphere model it goes to $A_0\beta_0$ (see eqn~\ref{eq: mihalas}). If we scale $\beta_0$ and $\tau_p$ by $\beta_0$ = $\tau_p/\tau_c = \tau_p/(\sigma_cN_H)$, and choose $f_c$ equal to $A_0$, the two curves of growth can be shifted on top of each other. 

\begin{figure}
 \centering
    {\includegraphics[width=\linewidth]{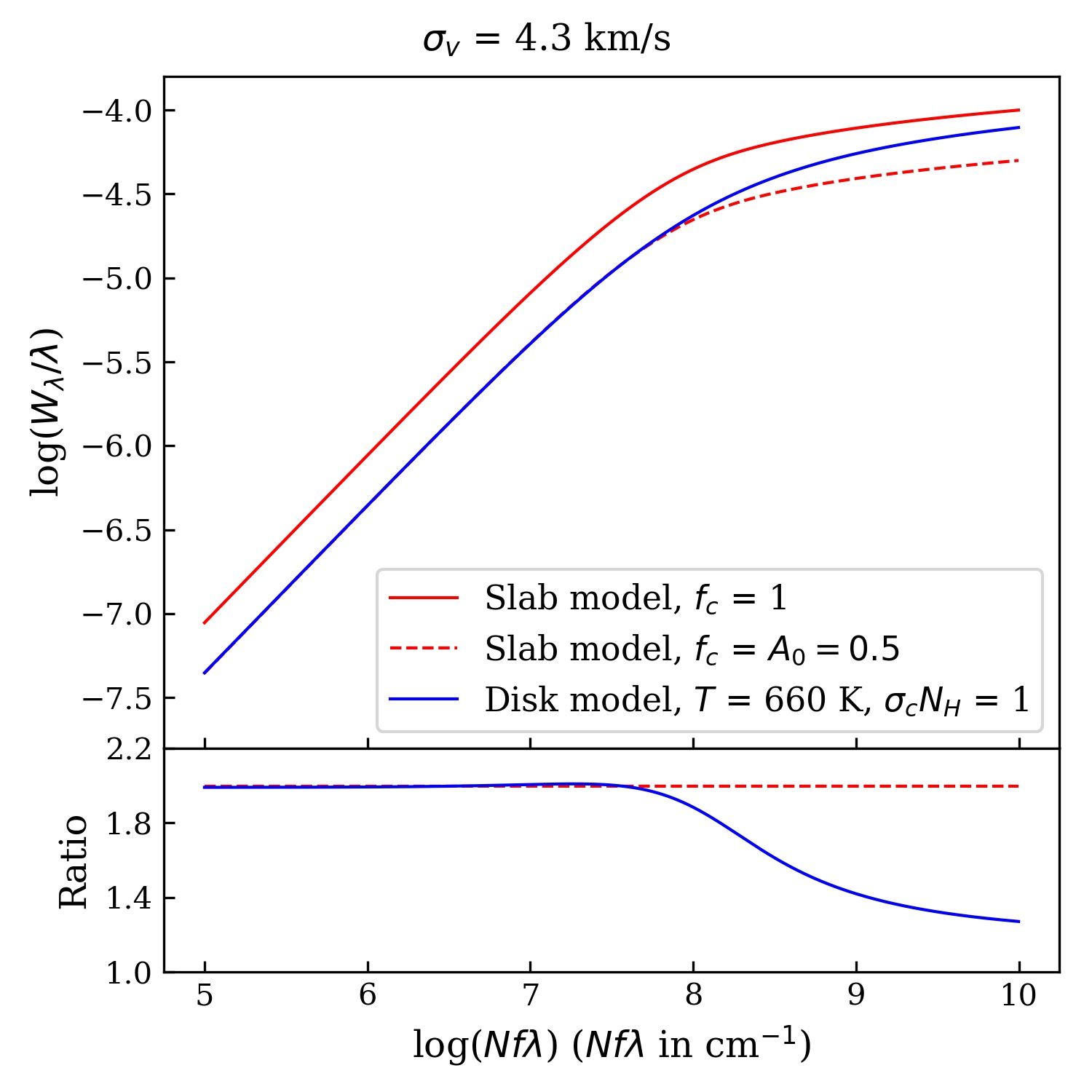}  }
    {\caption{\textit{Upper panel:} Curves of growth representing a slab model with and without a covering factor ($f_c = A_0$), and a stellar atmosphere model on a circumstellar disk adopting $\sigma_c N_H$=1. Both models were constructed with $\sigma_v = 4.3$~\kms, and $T = 660$~K adopted from the rotation diagram analysis of \thco\ is applied to the disk model. The central depth of an opaque line, $A_0$ under 660~K is 0.5. \textit{Lower panel:} the ratio of each
curve relative to that of the slab model without a covering
factor.}\label{fig:slab-at}}
    
\end{figure}

We construct the curves of growth from the two models above (eqn \ref{eq: draine} and \ref{eq: mihalas}) in $W_\lambda/\lambda$ versus $N_{l}f_{lu}\lambda$ in Figure~\ref{fig:slab-at}. As an example, the curve of growth in Figure~\ref{fig:slab-at} adopts $T = 660$~K and $\sigma_v$ = 4.3~\kms\ ($b = 6.1$~\kms) that are relevant for the \thco\ component of MIR2-H1 at $-38$~\kms (see \S~\ref{subsec:id}). Figure~\ref{fig:slab-at} also presents the slab model with and without a covering factor ($f_c = A_0$). We can formulate $\beta_0$ analogous to eqn ~\ref{drain-tau} by assuming that $\sigma_c N_H$ is 1 as for weak lines, {which essentially indicates that we see down to a continuum optical depth of unity}:
\begin{equation}
\beta_0 = \frac{1}{8\pi \sqrt{2\pi}} \frac{A_{ul} \lambda^2}{f_{lu} \sigma_v}\frac{g_u}{g_l}(N_l f_{lu} \lambda). \label{bn}
\end{equation}

Figure~\ref{fig:slab-at} illustrates how the two approaches are shifted on top of each other. The curves shift along the X-axis because of the difference between $\beta_0$ and $\tau_p$ and the curves shift along the Y-axis because of the $f_c$ versus $A_0$ factor. Specifically, when the two {curves of growth} overlap in the optically thin limit, $\tau_c = \sigma_c N_H = 1$. For large optical depth, lines in the slab model will saturate at $f_c$, while for the atmosphere, they go to $A_0$. However, even if we choose $f_c =A_0$, the approach to these limits is slightly different (Figure~\ref{fig:slab-at}).

In summary, for highly optically thick lines, the rotation diagram will severely underestimate the column density/abundance of the absorbing species, and a curve of growth approach is required. In a foreground cloud scenario, high optical depth transitions can be recognized by saturated line profiles with zero intensity. However, if the cloud only partially covers the continuum source, the line profile will not go to zero intensity even for highly optically thick lines. For absorption originating in the disk surface, a temperature gradient will naturally lead to non-zero intensity in the depth of the line. Introduction of an appropriate covering factor can make the two curve-of-growth approaches overlap and the two approaches show only subtle differences for modestly optically thick lines (Figure~\ref{fig:slab-at}).

\section{Results}\label{sec:results}

\begin{figure*}
\centering
    {\includegraphics[width=0.745\linewidth]{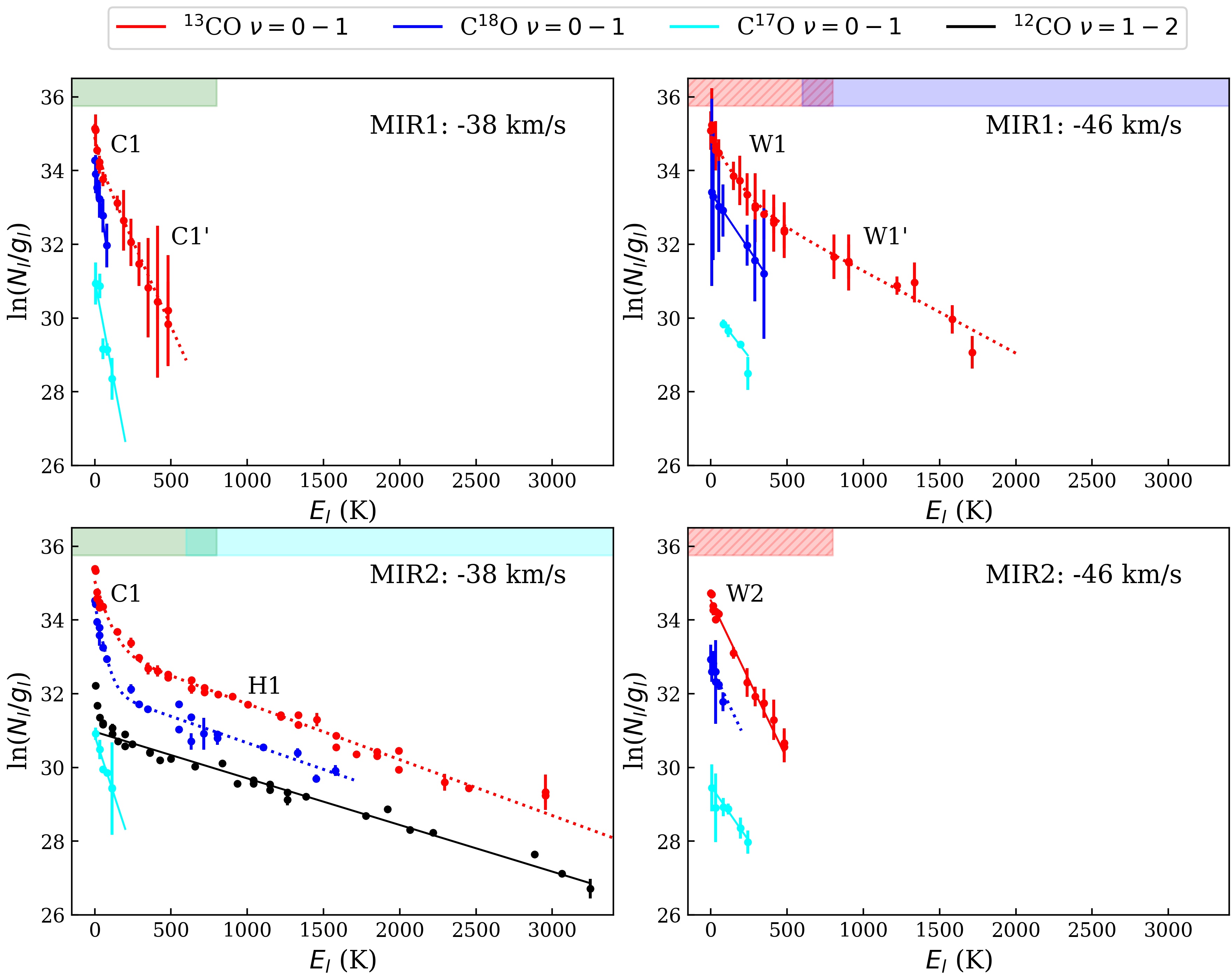}  }
    {\caption{Rotation diagrams of \thco~(\textit{red}), \ceio~(\textit{blue}), \cso~(\textit{cyan}), and \co\ \nuu=1--2~(\textit{black}) of a selected group of kinematic components. The colors of each component are consistent with those designated in Figure~\ref{fig:avg-fit} (see supplementary rotation diagrams of other identified components in Figure~\ref{fig:all-bd-app2}). 
    $N_l$ are derived from equation~\ref{eq:2} with Gaussian fitting. Solid lines represent fitting results of equation~\ref{eq:bd} and dotted lines are of equation~\ref{eq:bd2}. The derived \tex\ and $N_\textrm{tot}$ are listed in Table~\ref{table:a1a2}.}\label{fig:all-bd}}
\end{figure*}

As we have summarized the analysis methods in Section~\ref{sec:methods}, we present in this section the identification processes and the derived physical conditions of different components. We conduct the preliminary identification by iterating the decomposition of line profiles and the rotation diagrams in Section~\ref{subsec:id}, and present in Section~\ref{subsec:cog-slab} the procedures of modifications with the two curve of growth analyses. We discuss the properties of each identified component in Section~\ref{subsec:cog-star}.

\subsection{Optically Thin Slab Modelling}\label{subsec:id}

Because most velocity components in our data are blended, we first attempt to use multiple Gaussians to fit and decompose the non-saturated absorption profiles, assuming that the line is optically thin and its profile only consists of a Doppler core.  We derive the physical conditions of each identified kinematic component via rotation diagrams in Figure~\ref{fig:all-bd} (and in Fig.~\ref{fig:all-bd-app2} for supplementary plots). Assuming a slab model in LTE, we get $\tau$ and $N_l$ of each absorption line (see Table~\ref{tab:appex1} in the Appendix) with equation~\ref{eq:bd} and \ref{eq:bd2}, depending on whether the ln($N_l/(2J+1)$)-$E_l/k_B$ relation on the rotation diagram has a constant gradient or not.  Ideally, the identified kinematic components seen in the different isotopes with the same velocity center should originate from the same physical component, and have consistent properties such as line width and temperature.

\begin{figure*}
\centering
    {\includegraphics[width=0.99\linewidth]{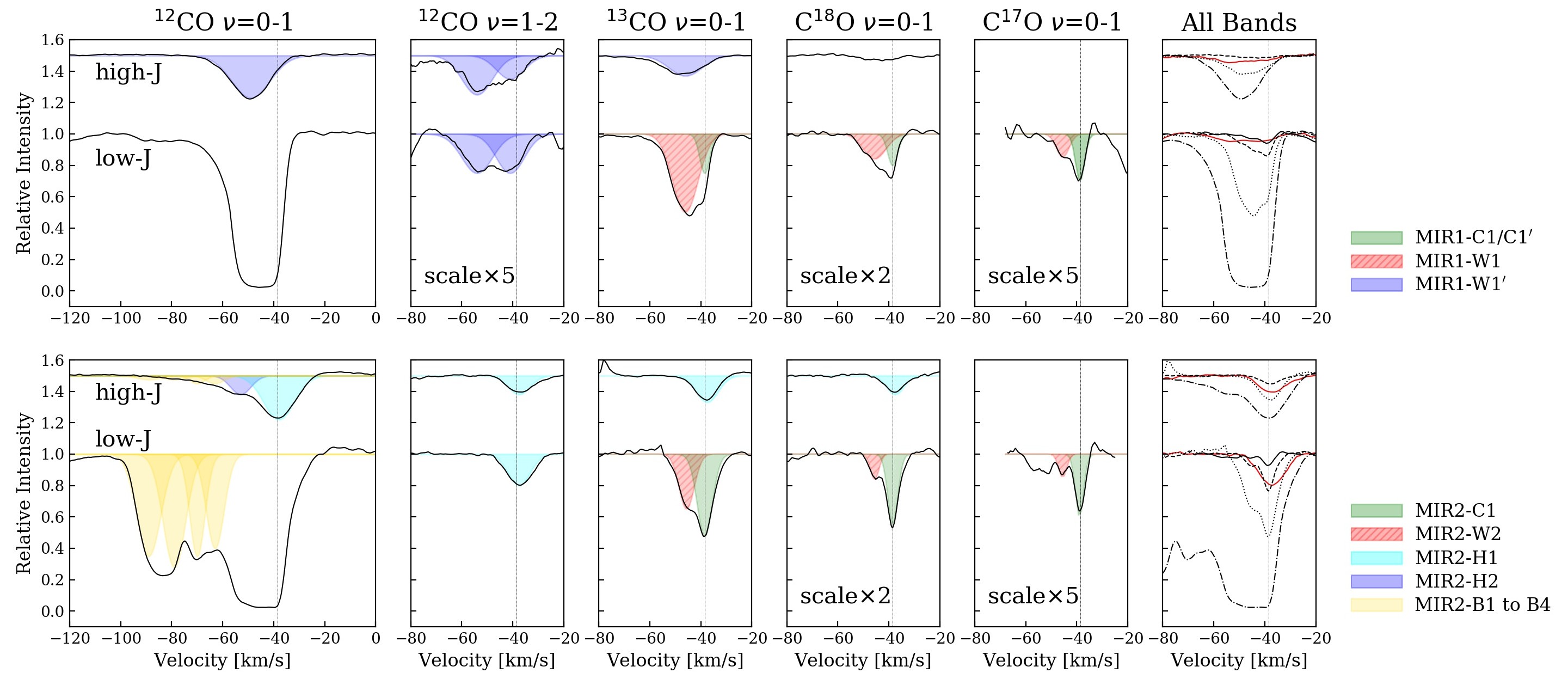}  }
    {\caption{Identified kinematic components of MIR1 and MIR2 on the average spectra of low- and high-energy transitions. The detailed line information, as well as the derived physical properties of each kinematic component, are listed in Table~\ref{table:a1a2}. ``C", ``W", ``H" stand for ``cold", ``warm", ``hot" and ``C1/C1$^\prime$'',  ``W1/W1$^\prime$'' indicate a change of temperature shown on rotation diagrams.  No components in the saturated part of \co\ spectra in low-$J$ are identified because we are not able to decompose distinct components there. Panels in the rightmost column summarize the identified components in the averaged spectra of all bands. The vertical dashed lines represent $v_{sys}$=-38.5~\kms. }\label{fig:avg-fit}}
\end{figure*}

We present the identified kinematic components in Figure~\ref{fig:avg-fit}, and list line properties and derived physical conditions in Table~\ref{table:a1a2}. We grouped and named kinematic components from different species at consistent velocities with similar velocity widths based on their temperatures. In those names, ``C", ``W", and ``H" stand for ``cold", ``warm", and ``hot". ``B" represents the high-velocity components that appear exclusively in \co\ \nuu=0--1 transition in MIR2 (B stands for bullets, see \S~\ref{subsec:bb}). For components of MIR1 at $-38$~\kms~(MIR1-C1/C1$^\prime$) and $-46$~\kms~(MIR1-W1/W1$^\prime$), and of MIR2 at $-38$~\kms~(MIR2-C1/H2), we see slope variation on rotation diagrams. We address below in Section~\ref{subsec:cog-star} whether this is due to an optical depth effect or a real temperature variation between different physical components.  

Properties of distinctive components among different species in  Table~\ref{table:a1a2} can not be easily reconciled. First, for all components with measurable \mbox{\nuu=0--1} and \nuu=1--2 lines (MIR1-W1$^\prime$, MIR2-H1, \mbox{MIR2-H2}), their temperatures derived from \co\ are much higher than that derived from isotopes. Second, the measured relative abundance ratios of isotopes are usually much smaller than the value found in local ISM (see Table~\ref{table:colratio}). Third, the velocity widths in different species do not always match. All of these issues reflect that optical depth effects are important. We justify below in Section~\ref{subsec:cog-star} that for some components, those inconsistencies can be reconciled by introducing a curve of
growth analysis, covering factors, and absorption in a disk photosphere.

\begin{deluxetable*}{l c rrccrr }
\tablecolumns{8}
\tabletypesize{\scriptsize}
\tablecaption{Physical Conditions of Components in Fig.~\ref{fig:all-bd} Derived Directly from Gaussian Fitted Profiles and Rotation Diagrams.  \label{table:a1a2}}
\tablehead{\colhead{Component} &  \colhead{Transitions} & \colhead{$E_l$} & \colhead{$J$} & \colhead{$v_\textrm{LSR}$} & \colhead{$\sigma_v$} & \colhead{$T_\textrm{ex}$} & \colhead{$N_\textrm{tot}$} \\
 &&\colhead{(K)} &&(\kms) &  (\kms)& \colhead{(K)}    &  \colhead{($\times 10^{16}$\percmsq)} \\
 (1) & (2) & \colhead{(3)} & \colhead{(4)} & \colhead{(5)} & \colhead{(6)} & \colhead{(7)} & \colhead{(8)}}
\startdata
 \multicolumn{8}{c}{MIR1} \\
  \hline
MIR1-C1 & \thco\ \nuu=0--1 &0--53 & 0--4 & -38.5 & 2.0 &  35.7$^{+16.8}_{-16.9}$ & 2.5$^{+1.5}_{-1.0}$  \\
&\ceio\ \nuu=0--1 & 0--79 & 0--5& -39.0 & 1.8 & 49.1$^{+33.2}_{-15.9}$  &1.3$^{+0.6}_{-0.3}$ \\
& \cso\ \nuu=0--1& 5--113 & 1--6 & -39.0 & 1.5  &  45.8$^{+24.7}_{-13.3}$ &  0.31$^{+0.11}_{-0.07}$\\
MIR1-C1$^{\prime}$  & \thco\ \nuu=0--1 & 148--481 & 7--13 & -38.8 & 2.0 & 120.2$^{+18.5}_{-17.0}$ & 5.5$^{+1.6}_{-1.5}$ \\
\hline
MIR1-W1 & \thco\ \nuu=0--1 &0--481& 0--13& -46.0 & 5.0 & 103.2$^{+43.7}_{-43.4}$ & 9.0$^{+4.6}_{-4.0}$ \\
&\ceio\ \nuu=0--1 & 0--553&  0--14 & -46.0 & 4.0  & 163.6$\pm$4.1 & 2.1$\pm$0.1 \\
& \cso\ \nuu=0--1& 81--243 & 5--9 &  -46.0 & 2.5 & 188.3$\pm$36.4& 0.6$\pm$0.1\\ 
\hline
MIR1-W1$^{\prime}$  & \thco\ \nuu=0--1  & 719--1994 & 16--27& -46.0 & 6.5 & 448.5$^{+86.5}_{-70.0}$ & 12.1$^{+4.1}_{-3.2}$\\
& \co\ \nuu=0--1 & 1937--5201 & 26--43 & -49.0 & 7.5 & 956.3$\pm$56.7 & 74.2$\pm$15.4\\
& \co\ \nuu=1--2 & 3100--4347 & 2--21 & -40.0 & 5.5 & 860.2$\pm$159.7 & 1.2$\pm$0.2\\   
&& 3100--4347 & 2--21 & -54.0 & 6.0 & 827.3$\pm$245.4& 1.4$\pm$0.4\\   
[0.5ex]
\hline
 \multicolumn{8}{c}{MIR2} \\
  \hline
 MIR2-C1 & \thco\ \nuu=0--1 &0--148& 0--7& -38.5 & 2.8 & 31.4$\pm$ 5.2 & 5.4$\pm$ 0.7 \\
&\ceio\ \nuu=0--1 & 0--79& 0--5 & -38.5 & 2.0 & 45.0$\pm$ 3.2 & 1.7$\pm$ 0.2\\
& \cso\ \nuu=0--1& 5-113 & 1--6 & -38.5 & 1.6 & 79.2$^{+13.8}_{-15.4}$ &  0.44$^{+0.07}_{-0.06}$ \\
\hline
MIR2-W2 & \thco\ \nuu=0--1 & 0--481 & 0--13 & -45.5 & 3.0 & 116.1$\pm$6.9 & 8.8$\pm$0.5\\
&\ceio\ \nuu=0--1 & 0--79 & 0--5& -45.5 & 1.7 & 97$^{+33.4}_{-32.9}$ & 0.7$^{+0.2}_{-0.2}$\\
& \cso\ \nuu=0--1& 31--243 & 1--9 & -45.5 & 1.6& 168.0$\pm$19.0& 0.24$\pm$\text{0.02} \\
\hline
MIR2-H1  & \thco\ \nuu=0--1 & 634--2956 & 13--33& -37.5 & 4.3 & 659.5$^{+35.2}_{-40.8}$ & 13.8$^{+0.9}_{-0.8}$\\
&\ceio\ \nuu=0--1 & 237--1578 & 9--24 & -37.5 & 3.4 & 695.9$^{+115.4}_{-89.3}$ & 2.3$^{+0.2}_{-0.2}$ \\
& \co\ \nuu=0--1 & 1937--5687 &  26--45 & -38.0 & 6.0 & 1162.1$\pm$55.1 & 65.3$\pm$8.6 \\
& \co\ \nuu=1--2 & 3088--6331& 1--33& -37.5 & 4.3 & 790.9$\pm$45.7 & 6.0$\pm$0.5 \\
\hline
MIR2-H2 & \thco\ \nuu=0--1 & 808--1336 &  17--22& -52.0 & 4.0 & 484.9$\pm$98.3 & 2.4$\pm$1.2\\
& \co\ \nuu=0--1 & 1937--4293 & 26--45 & -53.0 & 3.9 & 979.7$\pm$115.6 & 17.0$\pm$5.6\\
\hline
MIR2-B1 & \co\ \nuu=0--1 & 0--1275 & 0--21 & -89.0 & 3.5& 265.2$\pm$11.5 & 24.9$\pm$1.6 \\
MIR2-B2 & \co\ \nuu=0--1 & 0--2085 & 0--27  & -79.0 & 3.0  & 245.6$\pm$7.8& 22.8$\pm$1.1\\
MIR2-B3 & \co\ \nuu=0--1 & 0--2085 & 0--27  & -70.0 & 3.0  & 229.8$\pm$6.2& 13.6$\pm$0.6\\
MIR2-B4 & \co\ \nuu=0--1 & 0--2564 & 0--30  & -63.0 & 4.5& 351.7$\pm$12.9 & 18.2$\pm$1.0\\
\enddata
 \tablecomments{\,\, Column (1): Identified components. `C1$^{\prime}$' and `W1$^{\prime}$' represent the temperature gradient seen in component `C1' and `W1'. \\
 (5) Velocity of the line center.\\ (6) $\sigma_v$: the standard deviation of the Gaussian core, and equals to $b/\sqrt{2}$.\\ (7) \& (8): Derived temperatures and total column densities. Values with asymmetrical uncertainties were the 16th and 84th percentiles derived from MCMC when there is a temperature gradient seen in the rotation diagram (dashed lines in Fig.~\ref{fig:all-bd}).} 
\end{deluxetable*}

\begin{deluxetable}{c c cc }
\tablecolumns{4}
\tabletypesize{\scriptsize}
\tablecaption{Column Density Ratios Derived from Rotation Diagram Analysis \label{table:colratio}}
\tablehead{\colhead{Component} &  $N_{^{12}CO}$/$N_{^{13}CO}$ &  $N_{^{13}CO}$/$N_{C^{18}O}$ & $N_{C^{18}O}$/$N_{C^{17}O}$}
\startdata
Galactic Ratios & 66$\pm4$\tablenotemark{a} 
& 9.1$^{+3.7}_{-3.3}$\tablenotemark{b}& 4.16$\pm0.09$\tablenotemark{c}\\
MIR1-C1 & -- & 1.9$_{-1.1}^{+2.1}$ & 4.2$_{-1.8}^{+3.7}$ \\
MIR1-W1 & -- & 4.3$_{-2.0}^{+2.5}$ & 3.5$_{-0.6}^{+0.9}$\\
MIR1-W1$^{\prime}$ & 6.1$_{-2.3}^{+3.9}$ & -- & --\\
MIR2-C1 & -- & 3.2$_{-0.7}^{+0.9}$ & 3.9$_{-0.9}^{+1.1}$ \\
MIR2-W2 & -- & 12.6$_{-3.3}^{+6.0}$ & 2.9$_{-1.0}^{+1.2}$\\
MIR2-H1 & 4.7$_{-0.9}^{+1.0}$& 6.0$_{-0.8}^{+1.0}$ & -- \\
MIR2-H2 & 7.1$_{-3.9}^{+11.8}$& --& -- \\
[0.5ex] 
\enddata
 \tablenotetext{a}{ [$^{12}$C/$^{13}$C] of W3(OH) measured in \citep{milam05}.}
 \tablenotetext{b}{  [$^{16}$O/$^{18}$O] = (58.8 $\pm$ 11.8) $\times D_\textrm{GD}+(37.1\pm82.6)$ \citep{wilson94}, which is 601.6$\pm$195.9 for \wt. We adopt $N$(\thco)/$N$(\ceio) = [\co/\ceio]/[\co/\thco] = [$^{16}$O/$^{18}$O]/[$^{12}$C/$^{13}$C] in the table.}
 \tablenotetext{c}{\citet{wouterloot08}.}
\end{deluxetable}

\subsection{Two Curve of Growth Analyses}\label{subsec:cog-slab}

We discuss in this section the detailed analysis procedure for all identified components in Table~\ref{table:a1a2}. MIR2-W2 is not included, because its $N$(\thco)/$N$(\ceio) is even greater than the galactic [\thco/\ceio] value. It is likely that there is an unresolved \thco\ component of which the \ceio\ component is buried in the noise, as indicated by the much broader \thco\ width in Table~\ref{table:a1a2}.

\subsubsection{Slab Model of a Foreground Cloud}

\begin{figure*}[!t]
    \centering
    \includegraphics[width=\linewidth]{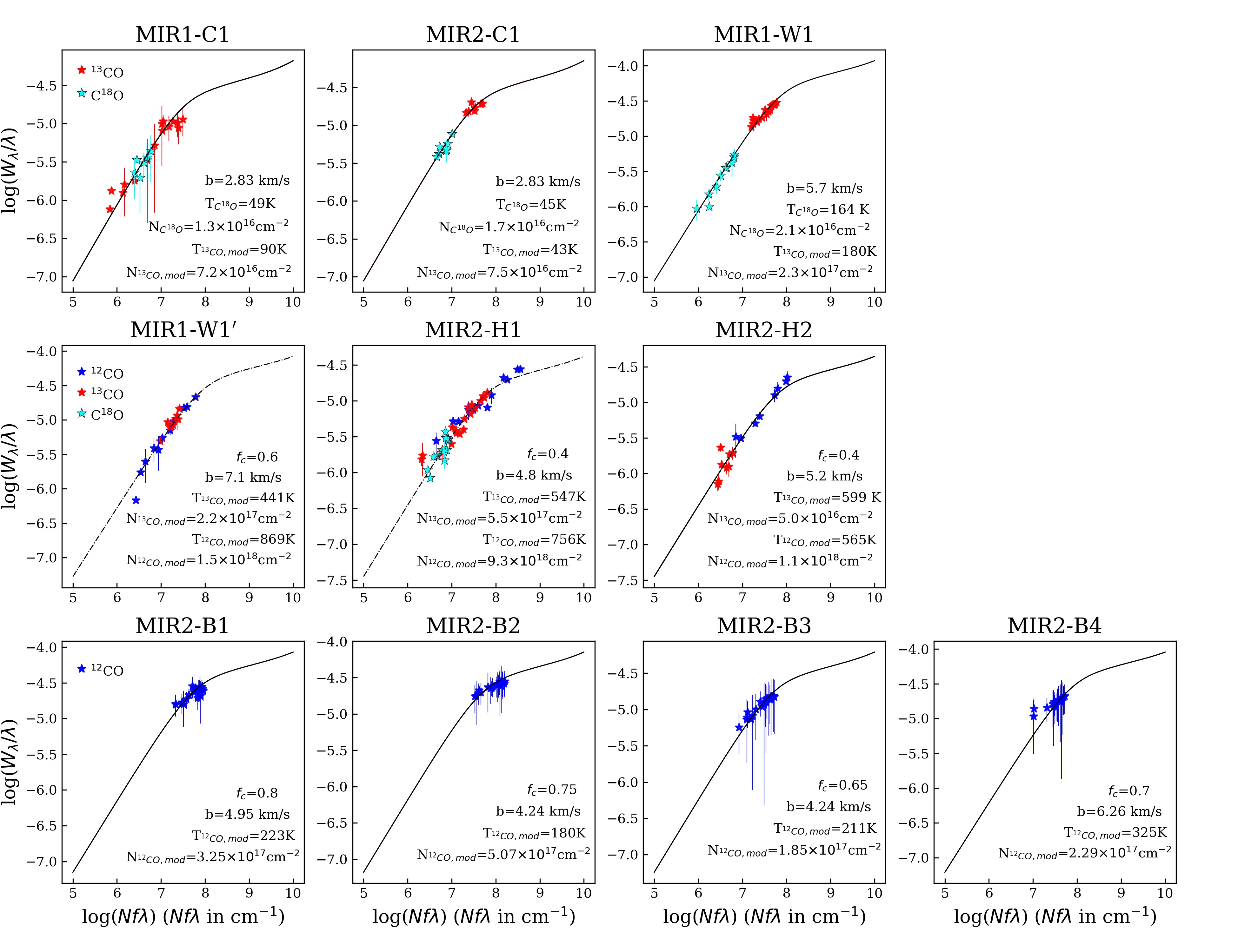}
    \caption{The best-fitting results of the observed log$_{10}$($W_\lambda/\lambda$) versus log$_{10}$($Nf\lambda$) and the theoretical curve of growth of a slab model on each individual component. The theoretical curve of growth and the values of $Nf\lambda$ are calculated based on the best-fitted column density and temperature. Dashed curves of growth plotted on MIR1-W$^\prime$ and MIR2-H1 indicate certain problems in the fitting results (\S~\ref{subsec:slab-w1} and \S~\ref{subsec:slab-h1h2}).}
    \label{fig:slab}
\end{figure*}

\begin{figure*}[!t]
    \centering
    \includegraphics[width=0.89\linewidth]{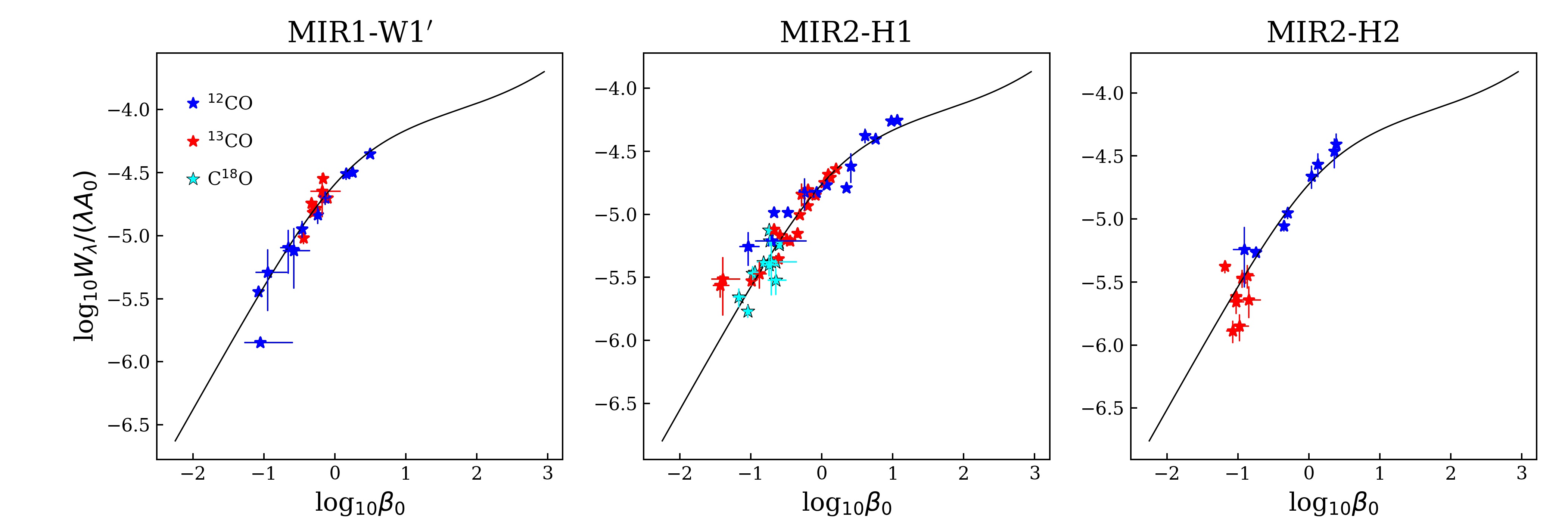}
    \caption{The best-fitting results of log$_{10}$($W_\lambda/(\lambda A_0)$)-log$_{10}$($\beta_0$) and the theoretical curve of growth of a stellar atmosphere model on MIR1-W1$^\prime$, MIR2-H1, MIR2-H2.  The equivalent width in velocity space, $W_\lambda$ (= $\lambda W_v/c$) of each molecular dataset are observable, and the Doppler width, $b$, is adopted from the smallest line width observed in \thco\ or \ceio~(see texts in Section~\ref{subsec422}). The theoretical curve of growth, the values of $\beta_0$ and $A_0$ are calculated based on the best fitted $N_{tot}/N_H$ and temperature.}
    \label{fig:disk-slab-results}
\end{figure*}

Assuming a foreground slab model, we use the curve of growth analysis to account for optical depth effects and to determine the column density, temperature, and covering factor of a kinematic component of bands from all relevant CO isotopes. When a component is observed in multiple species, the smallest line width observed in \co, \thco, or \ceio\ is used to estimate the Doppler parameter. We do not use the line widths observed in \cso\ lines because rather large uncertainties would be introduced given the too few data points and poor baseline fitting. When a partial covering factor is necessary, it is bounded by the upper limit of the absorption intensity in the data sets. We obtain ($T_\textrm{ex}$, $N_\textrm{tot}$) by looking for the smallest reduced $\chi^2$ in fitting the observable $W_\lambda/\lambda$ to the curve of growth (equation~\ref{eq: draine}), and estimate the 1$\sigma$ uncertainty by looking for the $\chi^2_{{r},\,{min}}$+$\Delta/dof$ contour in the ($T_\textrm{ex}$, $N_\textrm{tot}$) grid, where $\Delta$ is the $\chi^2$ critical value for a significance level of 68.3$\%$ and $dof$ is degree of freedom, $n-2$ ($n$ is the sample size). We summarize the results in Figure~\ref{fig:slab} and Table~\ref{table:cor-slab}.

\subsubsection{Stellar Atmosphere Model of a Disk}\label{subsec422}

Section~\ref{subsec:id} reveals three hot components (MIR1-W1$^\prime$, MIR2-H1, and MIR2-H2) with temperatures between 500--700~K. This range is close to the dust temperature required to produce the observed continuum emission of the mid-IR disks. We therefore consider the possibility that three components are present in the photosphere of the disk and are absorbing against the continuum there. Similar to the curve of growth analysis on a foreground slab model, we apply the grid search method by fitting the observable $W_v/c$ that equals to $W_\lambda/\lambda$ to the curve of growth (eqn~\ref{eq: mihalas}) assuming pure absorption (see \S~\ref{mode:sam}). In the fitting procedure, we also reconcile the properties of different species by fitting with a Doppler width that is the smallest width measured among different species, i.e. \thco\ for MIR1-W1$^\prime$ and MIR2-H2, and \ceio\ for MIR2-H1. We present the fitting results in Figure~\ref{fig:disk-slab-results} and Table~\ref{table:cor-disk-star}. We note that since the exact value of $\sigma_c$ influences the derived absolute abundance ratio, we only present the relative abundance of different species in the same component in Table~\ref{table:cor-disk-star}. If we assume that \co\ has a constant relative abundance of $1.6\times10^{-4}$  \citep{cardelli96, sofia97}, we may derive $\sigma_c$ as listed in Table~\ref{table:sig_c}. The variation of $\sigma_c$ for about an order of magnitude may convey information on the dust aggregation characteristics, for example, the dominant size in the aggregation distribution \citep{ormel11}. 

\subsection{Properties of Individual Components}\label{subsec:cog-star}

\begin{deluxetable*}{cc ccccc cc}[!t]
\tablecolumns{9}
\tabletypesize{\scriptsize}
\tablecaption{Results for the Slab Model, using a Curve of Growth Analysis   \label{table:cor-slab}}
\tablehead{\colhead{Component} & $f_c$ &$\sigma_v$ & $T_{^{12}\textrm{CO, } \textrm{mod}}$  & $T_{^{13}\textrm{CO, } \textrm{mod}}$   & $N_{^{12}\textrm{CO, } \textrm{mod}}$ & $N_{^{13}\textrm{CO, } \textrm{mod}}$ & $N_{^{12}\textrm{CO/}^{13}\textrm{CO, } \textrm{mod}}$& $N_{^{13}\textrm{CO/C}^{18}\textrm{O, } \textrm{mod}}$ \\  &  & (\kms)&(K) &(K)    &  ($\times 10^{17}$\percmsq) & ($\times 10^{16}$\percmsq) 
}
\startdata
MIR1-C1 & 1.0 & 1.8 &--& 90$_{-27}^{+70}$ & -- & 7.2$_{-3.0}^{+6.4}$&--& 5.4$_{-3.3}^{+8.1}$\\
MIR2-C1 & 1.0 & 2.0 &--& 43$_{-13}^{+27}$ & -- & 7.5$_{-1.6}^{+3.7}$& -- & 4.4$_{-1.3}^{+3.1}$\\
MIR1-W1 & 1.0 & 4.0 &--& 180$_{-14}^{+11}$ & -- & 23.0$_{-1.1}^{+2.5}$ & -- & 10.9$_{-1.0}^{+1.8}$ \\
\hline
MIR1-W1$^\prime$ & 0.6 & 5.0  & 869$_{-131}^{+155}$ & 441$_{-65}^{+94}$ & 14.8$_{-3.8}^{+6.1}$  & 21.7$_{-2.2}^{+4.7}$ & 6.8$_{-2.7}^{+3.9}$ & --\\
MIR2-H1  & 0.4 & 3.4 &  756$_{-45}^{+49}$ & 547$_{-37}^{+44}$ &  92.5$_{-20.5}^{+27.7}$  & 55.0$_{-6.1}^{+4.9}$ & 16.8$_{-4.8}^{+7.8}$ & 9.6$_{-1.1}^{+0.9}$ \\
MIR2-H2  & 0.4 & 3.7  & 565$_{-92}^{+122}$ & 599$_{-186}^{+518}$ & 11.0$_{-3.9}^{+9.2}$  & 5.0$_{-1.4}^{+2.6}$  & 22.0$_{-11.5}^{+36.6}$ & --\\
\hline
MIR2-B1  &0.8 & 3.5 & 223   &--&3.25 & --& --& --\\
MIR2-B2  &0.75 & 3.0 & 180 & --&5.07 & --& --& --\\
MIR2-B3  &0.65& 3.0 & 211&--&1.85 & --& --& --\\
MIR2-B4  & 0.7 & 4.5 & 325 & -- &2.29 & --& --& --\\[0.5ex] 
\enddata
\tablecomments{For MIR2-B1 to B4, decomposing the blended line profiles may introduce large uncertainties. We hence do not report the uncertainty level in the derived physical conditions.}
\end{deluxetable*} 

\begin{deluxetable*}{ccc c cc c c}[!t]
\tablecolumns{8}
\tabletypesize{\scriptsize}
\tablecaption{Results for the Disk Atmosphere Model \label{table:cor-disk-star}}
\tablehead{\colhead{Component} & $\sigma_v$ & $T_{^{12}\textrm{CO, } \textrm{mod}}$   &  $T_{^{13}\textrm{CO, } \textrm{mod}}$ &   $T_{\textrm{C}^{18}\textrm{O, } \textrm{ex}}$ &  $X$[$^{12}\textrm{CO}$]/$X$[$^{13}\textrm{CO}$] & $X$[$^{13}\textrm{CO}$]/$X$[$\textrm{C}^{18}\textrm{O}$] \\   &   (\kms)&(K)     & (K)    & (K) & &
}
\startdata
MIR1-W1$^\prime$ & 5.0 & 709$_{-101}^{+136}$ & 474$_{-96}^{+152}$  & -- & 17.1$_{-5.9}^{+9.3}$ & --\\
MIR2-H1  & 3.4 & 662$_{-28}^{+33}$  & 507$_{-37}^{+47}$ & 676$_{-128}^{+230}$ & 41.1$_{-9.8}^{+11.8}$ & 11.4$_{-3.3}^{+3.4}$  \\
MIR2-H2  & 3.7 &  482$_{-70}^{+97}$ &  542$_{-201}^{+795}$ & -- & 24.6$_{-16.4}^{+36.1}$ \\[0.5ex] 
\enddata
\tablecomments{The absolute abundance of a species, $X$[mol], is defined as $N_{\textrm{mol}}/N_H$. Values of $X$[mol] in this table are dependent on the chosen dust opacity, therefore we only report the relative abundance of different species in the table.}
 \end{deluxetable*}

 \begin{deluxetable}{ccc}[!t]
\tablecolumns{3}
\tabletypesize{\scriptsize}
\tablecaption{Values of $\sigma_c$ Derived from the Curve of Growth Analysis in the Disk Model   \label{table:sig_c}}
\tablehead{\colhead{Component} & \colhead{$\sigma_c$} (cm$^2$/H-nucleus)& \colhead{$N$(\co) (cm$^{-2}$)}\\ (1) &(2) &(3)}
\startdata
MIR1-W1$^\prime$ & 2.7$_{-0.8}^{+1.0} \times 10^{-23}$ & 5.9$_{-1.6}^{+2.5} \times 10^{18}$\\
MIR2-H1  & 6.6$_{-1.2}^{+1.3} \times 10^{-24}$ & 2.4$_{-0.5}^{+0.5} \times 10^{19}$ \\
MIR2-H2  & 1.2$_{-0.5}^{+0.8} \times 10^{-22}$ & 1.4$_{-0.6}^{+1.2} \times 10^{18}$ \\[0.5ex] 
\enddata
\tablecomments{Column (2): Assuming a constant $X$[\co] of $1.6\times10^{-4}$, we derive values of $\sigma_c$ via  $(N_{\textrm{mol}}/N_H)(\sigma_c/\sigma_{c0})$ = $1.6\times10^{-4}$, in which $\sigma_{c0}$ = {7$\times 10^{-23}$ cm$^2$/H-nucleus following \citet{barr20}}.\\ Column (3): Assuming that we are looking at the column density depth where the dust opacity approaches 1 (and equivalently, $\sigma_c N_\textrm{H} = 1$), $N_{\textrm{mol}} = 1.6\times 10^{-4}/\sigma_c$.}
\end{deluxetable} 
 
\subsubsection{MIR1-C1 and MIR2-C1}

\begin{deluxetable}{r c c cc }[!t]
\tablecolumns{5}
\tabletypesize{\scriptsize}
\tablecaption{Parameters of the Toy Model \label{table:toy} for MIR1-W1$^\prime$}
\tablehead{\colhead{Component} & $f_c$ & $b$ (\kms) & T (K) & $N$(\co) (\percmsq) }
\startdata
MIR1-W1$^\prime$-B & 0.2 &  5.5 & 449 & 3.3$\times 10^{19}$ \\
MIR1-W1$^\prime$-N & 0.5 & 4.0 & 449 & 2.7$\times 10^{18}$ \\
\enddata
\tablecomments{Assuming $N$(\co)/$N$(\thco)=66 \citep{milam05}.}
\end{deluxetable} 

MIR1-C1 and MIR2-C1 are the two narrow low-$J$ components ($<$~3~\kms) detected in \thco, \ceio, \cso\ at -38~\kms\ sharing similar physical conditions. Considering the results from the rotation diagram analysis (Table~\ref{table:a1a2}), for MIR1-C1, the \thco/\ceio\ column density ratio, 1.9$^{+2.1}_{-1.1}$ is much less than the expected value of 9.1 (Table~\ref{table:colratio}). Besides, the lines are not fit with a single temperature: the high-$J$ levels reveal the presence of CO gas with a much higher excitation temperature. Hence, optical depth effects are indicated. The curve of growth analysis reconciles the temperatures of \thco\ and \ceio. As we present in Table~\ref{table:cor-slab}, for MIR1-C1/C1$'$, a temperature of 90 K resolves the temperature difference for the C1/C1$'$ components in the \thco\ data. However, the \ceio\ excitation temperature is still discrepant (49~K, Table~\ref{table:a1a2}). For MIR1-C1/C1$'$, the isotopic column density ratios agree within the uncertainty level. 

For MIR2-C1, although the column density ratio is sufficiently uncertain that they could be in agreement, the excitation temperatures of these two isotopologues differ. Hence, here too, optical depth effects might be important. Taking these into account, the temperature becomes 43 K but the isotopologue abundance ratio, 4.4$^{+3.1}_{-1.3}$, remains low compared to the expected ratio in the ISM (Table~\ref{table:cor-slab}). 

\subsubsection{MIR1-W1 and MIR1-W1$^{\prime}$}\label{subsec:slab-w1}

MIR1-W1 and MIR1-W1$^{\prime}$ are the two components at $\sim-60$ to $-40$~\kms. Their difference in temperature is indicated by the slope variation seen in the rotation diagram. MIR1-W1 is the cooler component. Although the rotation diagram analysis results in a much lower temperature of \thco~(103~K) than that of \ceio~(164~K) and a $N$(\thco)/$N$(\ceio) of only 4.3$_{-2.0}^{+2.5}$ (Table~\ref{table:colratio}), we can reconcile the properties of the two species by adopting the Doppler width of \ceio\ to \thco\ with the curve of growth analysis. The modified temperature of \thco\ is 180$_{-14}^{+11}$, and the modified relative column density is 10.9$_{-1.0}^{+1.8}$ (Table~\ref{table:cor-slab}). 

Properties of the warmer component MIR1-W1$^{\prime}$ are more complicated. Firstly, the line profiles in different transition bands are not consistent. As Table~\ref{table:a1a2} shows, both low- and high-J lines in \thco\ center at $-46$~\kms, while the centers of unsaturated high-J \co\ \nuu=0--1 lines are at $-49$~\kms. \co\ \nuu=1--2 lines have double peaks with one at $-40$~\kms\ and one at $-54$~\kms. The comparison is more clearly illustrated in the final plot on the first row of Figure~\ref{fig:avg-fit}, where the average spectra of all bands are overlaid. \co\ and \thco\ seem to share the red wing for high-J lines.

The complexity seen in the line profiles indicates that they arise in somewhat different kinematic components and hence we do not expect them to fall on a single rotation diagram or have an abundance ratio consistent with the isotope ratio. We present below the analysis over MIR1-W1 and MIR1-W1$^\prime$ for completeness.

With the rotation diagram analysis, the temperature of \thco~(449~K) is much less than that of \co~(956~K), and the relative column density ratio is only 6.1$_{-2.3}^{+3.9}$ (Table~\ref{table:colratio}). Adopting a Doppler width of 7.1~\kms\ and a fractional covering factor of $\sim$0.6, the saturated intensity, does not help to solve this problem. As we illustrate in Figure~\ref{fig:slab}, after the modification with $b$ and $f_c$, the \co\ and \thco\ are still located on the linear part of the curve of growth. Therefore, we cannot reconcile the properties of \co\ and \thco\ with a slab model assuming that \co\ and \thco\ lines each contain a single component. Considering substructures {can} work. For example, fixing the temperature of \co\ to that of \thco\ derived from the rotation diagram, and adopting a $N$(\co)/$N$(\thco) ratio of 66 \citep{milam05}, we may artificially fit the line profiles with a narrow component (MIR1-W1$^\prime$-N, $f_c$=0.2) dominating the line peak, and a broad component (MIR1-W1$^\prime$-B, $f_c$=0.5) dominating the wing (see Table~\ref{table:toy}).

Applying a stellar atmosphere model {can} unify the temperatures of \co\ and \thco. As we present in Figure~\ref{fig:disk-slab-results} and Table~\ref{table:cor-disk-star}, the dataset of \co\ moves to the logarithmic part of the curve of growth. The 1$\sigma$ temperature ranges of \co\ and \thco\ are also comparable. However, we were only able to increase $X$[$^{12}\textrm{CO}$]/$X$[$^{13}\textrm{CO}$] to 17.1$_{-5.9}^{+9.3}$. Adopting a smaller $\sigma_v$ increases the ratio further by at most up to twice, therefore it reinforces that \co\ and \thco\ do not originate from exactly the same component. 

\subsubsection{MIR2-H1}\label{subsec:slab-h1h2}

MIR2-H1 is close to $v_\textrm{sys}$ at $-38$~\kms\, the same velocity as MIR1-C1/MIR2-C1, but it is intrinsically different. MIR2-H1 appears in high-$J$ lines, and has a much broader line width ($\sigma_v$ = 4.3~\kms) than \mbox{MIR1-C1} and MIR2-C1 do. This component is hot enough to excite the vibrational band \co\ \nuu=1--2. If we assume that the total $N_0$, the column density in the $\nu$=0 level, for \co\ is 66 times that of \thco, and compare it with $N_1$, the total column derived from \co\ \nuu=1--2 in Table~\ref{table:a1a2}, we can derive a vibrational excitation temperature, $T_\textrm{vib}$ of 613~K via the Boltzmann's equation,
\begin{equation}
N_1/N_0  = \textrm{exp}(-3083.11/T_{\textrm{vib}}).  
\end{equation}
This vibrational excitation temperature is consistent with the rotational excitation temperature and firmly links the absorption in the 0--1 and 1--2 transitions.

While the temperatures of MIR2-H1 derived from the rotation diagram analysis of the \thco, \ceio, and \co\ \nuu=1--2 of the MIR2-H1 components are consistent with each other, they do not agree with the derived properties of the \co\ \nuu=0--1 component. The temperature of \co\ \nuu=0--1 derived from the rotation diagram is $\sim$~1000~K compared to 650–750 K for the isotopologues and vibrationally excited transitions. In addition, the derived $N$(\co)/$N$(\thco) is only 4.7$_{-0.9}^{+1.0}$ (Table~\ref{table:colratio}). Similar to the MIR1-W1$^\prime$ component, a slab model with $f_c$ = 1 is not correct. Otherwise, we would expect the absorption intensity of this component to approach zero in low-J lines, which is also not seen in our observed spectra.

We present in Figure~\ref{fig:slab} and {\ref{fig:disk-slab-results} the} curve of growth analysis of MIR2-H1 on a modified slab model ($f_c$=0.4, $\sigma_v$=4.8~\kms) and a stellar atmosphere model. As the \co\ \nuu=0--1 dataset moves to the logarithmic part with both analyses, we confirm that the \co\ \nuu=0--1 absorption profiles saturate in the Doppler core. {We consider that, when the temperatures  agree within 1$\sigma$, the different isotopologues probe the same gas (Table~\ref{table:cor-slab} and \ref{table:cor-disk-star}).} $N$(\co)/$N$(\thco) increases to 16.8$_{-4.8}^{+7.8}$ and 41.1$_{-9.8}^{+11.8}$, and $N$(\thco)/$N$(\ceio) increases to  9.6$_{-1.1}^{+0.9}$ and 11.4$_{-3.3}^{+3.4}$ (Table~\ref{table:cor-slab} and \ref{table:cor-disk-star}). For the slab model, the increased new column densities result in a lower $T_{vib}$ of \mbox{\co\ \nuu=1--2} of 513~K, suggesting that the particle density in the foreground cloud does not reach the critical density of the vibrational band. The vibrational level is likely subthermalized. 

\begin{figure*}[!t]
    \centering
    \includegraphics[width=\linewidth]{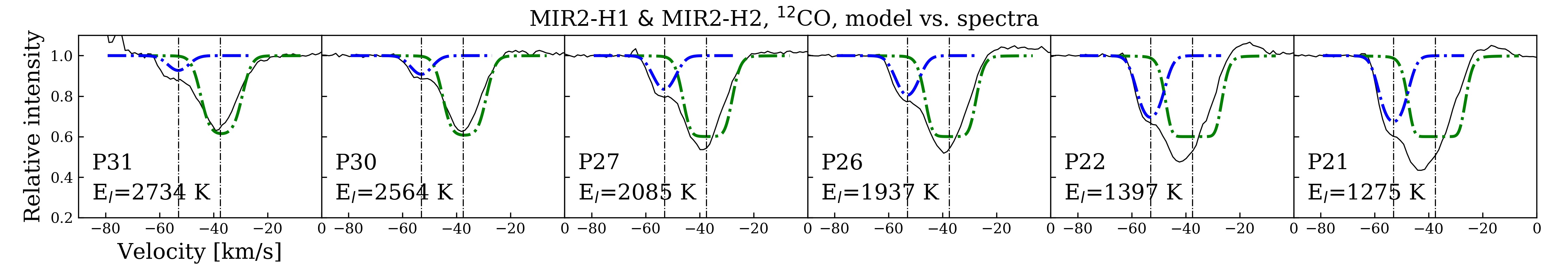}
    \caption{Comparison between the observed \co\ spectra (\textit{black}) and the modeled spectra (\textit{green and blue dashed}) under the curve of growth modification with a slab model on MIR2-H1 (Section~\ref{subsec:slab-h1h2}) and MIR2-H2 (Section~\ref{subsec:mir2h2}).}
    \label{fig:mir2-h1h2}
\end{figure*}

Although the equivalent widths fit nicely with the curve of growth, we found a mismatch between the profiles of the modeled spectra and the observed spectra for both models. This is illustrated in Figure~\ref{fig:mir2-h1h2} which takes the modified slab model as an example. For $J<$30 lines, the saturated modeled (green) spectra do not match the red wing of the observed spectra at $v_\textrm{sys}>-38$~\kms. We interpret this mismatch with an extra emission component in the red wing region, which is in emission. This component was also evident in the data of \citet[][component E]{mitchell91}.

\begin{figure*}
    \centering
    \includegraphics[width=\linewidth]{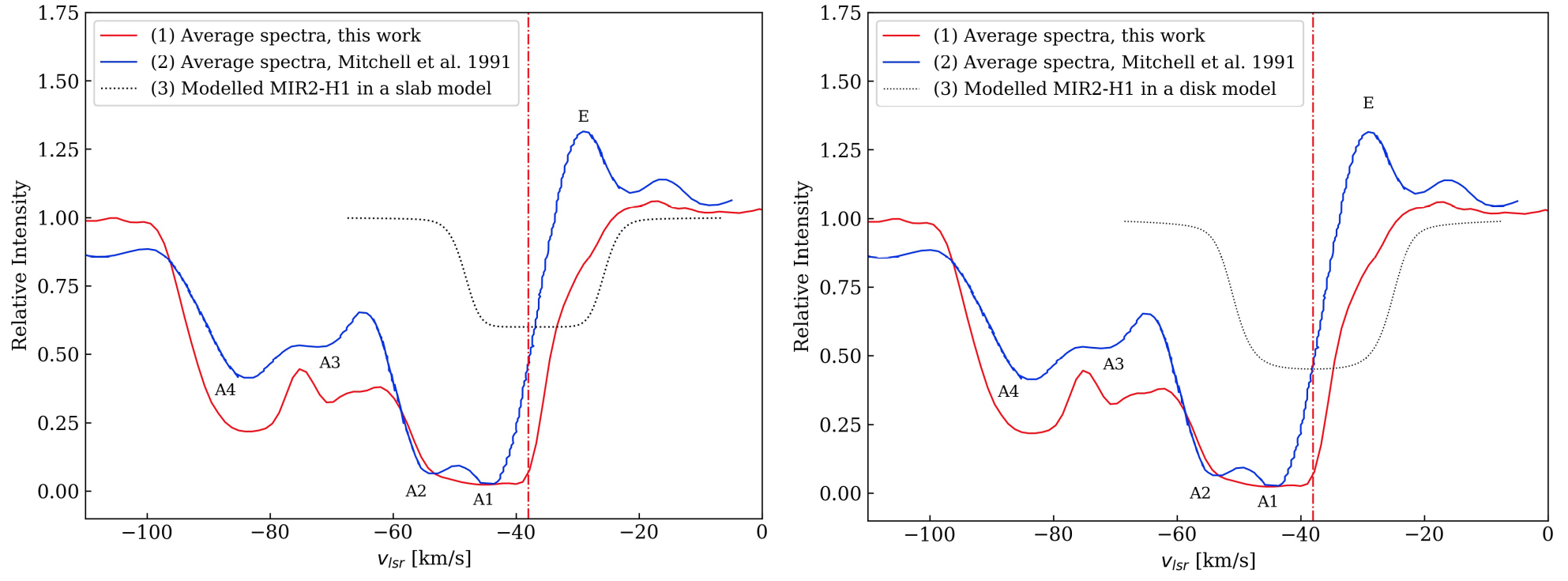}
    \caption{Comparison between the residual of the average of observed MIR2 spectra (P3, P6, P7, P8, P9, P12, R1, R3, and R7) minus that of the modeled MIR2-H1 spectra (\textit{left}: the slab model; \textit{right}: the disk model) and the average spectra reported in \citet{mitchell91}, in which MIR1 and MIR2 were not distinguished. Both panels show that the difference between the observed spectra and the model may be a correspondence of the emission component ``E" found in \citet{mitchell91}.}
    \label{fig:emission}
\end{figure*}

We plot the average modeled spectra of MIR2-H1, the observed spectra of MIR2, and the average spectra of \wt\ from \citet{mitchell91} in Figure~\ref{fig:emission}. All the spectra were of \co\ and were averaged over P3, P6, P7, P8, P9, P12, R1, R3, and R7 following \citet{mitchell91}, in which MIR1 and MIR2 were not distinguished, and a potential emission component ``E" at \mbox{$\sim-30$~\kms} was reported. Although the modeled MIR2-H1 in both models does not match the observed spectra at \mbox{$> -38$~\kms}, the residual between the models and the observed spectra indicates an emission component. Compared to the observed spectra in \citet{mitchell91}, this emission component may correspond to the ``E" component there. This emission component is visible up to $J$=22, suggesting that it is rather warm. If this component is real, the comparable intensity indicates a comparable ratio between the emission area relative to the observation fields, which is a 2.5\asec\ aperture covering the whole binary in \citet{mitchell91} and a 0.375\asec\ wide slit in front of MIR2 in this study. Further spatially resolved spectroscopy is required to confirm the reality of this emission component and its physical characteristics.

\subsubsection{MIR2-H2}\label{subsec:mir2h2}

MIR2-H2 is the other component observed in \co\ and \thco\ simultaneously. $N$(\co)/$N$(\thco) of 7.1$_{-3.9}^{+11.8}$ derived from the rotation diagram analysis (Table~\ref{table:colratio}) also indicates an underestimation of the \co\ column density. We reconcile the properties of \co\ and \thco\ by adopting $f_c$ = 0.4 and $b$ = 3.7~\kms~(Figure~\ref{fig:slab} and \ref{fig:mir2-h1h2}) 
in the slab model. We adopt $b$ = 3.7~\kms\ for the atmosphere model. For both models, the modified temperatures of \co\ and \thco\ are comparable. The corrected $N$(\co)/$N$(\thco) increases to  22.0$_{-11.5}^{+36.6}$ and 24.6$_{-16.4}^{+36.1}$ (Table~\ref{table:cor-slab} and \ref{table:cor-disk-star}), and the large uncertainties are due to the very few measurements of \thco\ lines on this component.

\subsubsection{MIR2-B1 to B4}\label{subsec:bb}

As iSHELL observations distinguish spectra originated from the binary separately, the absorption features of \co\ \nuu=0--1 between $\sim-60$ to $-90$~\kms\ were found to be exclusively associated with MIR2. No absorption lines of isotopologues were detected in this velocity range, indicating that these ``B" components have much lower column densities than those between $-38$ to $-60$~\kms. We decompose the blended profile by four Gaussians (MIR2-B1 to B4) and derive similar temperatures ($\sim$230--350~K) and total column densities ($\sim 2\times 10^{17}$~\percmsq) from rotation diagram analysis. 

For low-J lines of MIR2-B1 to B4, the steep slopes in the rotation diagrams together with the flap-top line profiles suggest line saturation in the Gaussian cores. We apply a covering factor of 0.8, 0.75, 0.65, and 0.7 based on the upper limit of the absorption intensities in the curve of growth analysis. We stress that there is a $\sim$5--7\% uncertainty due to contamination of the binaries in the spectral extraction process (Section~\ref{sec:obs}). As shown in Figure~\ref{fig:slab}, all the four components are on the logarithmic part. The corrected temperatures (180--325~K) have a minor decrease, and the column densities of each component are increased by less than a factor 2. We stress that our decomposition on the blended line profiles introduces quite large uncertainty in these estimates and therefore do not report them in Table~\ref{table:cor-slab}.

\section{Discussion}\label{section: discussion}

\begin{deluxetable*}{cccccccccccc}[!t]
\tablecolumns{12}
\tabletypesize{\scriptsize}
\tablecaption{Physical Conditions of Identified Components \label{table:sum}}
\tablehead{\colhead{Comp.} & \colhead{$v_\textrm{LSR}$} & \colhead{\tk} & \colhead{$f_c$} &  \colhead{$N_{\textrm{H}_2}$ } &  \colhead{$J_\textrm{max}$} & \colhead{log($n_\textrm{crit}$)} & \colhead{$d$ } & \colhead{$\sigma_\textrm{obs}$} & \colhead{$\sigma_v$}  & \colhead{Heating} & \colhead{Ref.} \\
& (\kms)&(K)& & ($\times 10^{22}$\percmsq)&& (\percmcu)& (AU) &(\kms) &  (\kms)& \\
(1) & (2) & \colhead{(3)} & \colhead{(4)} & \colhead{(5)} & \colhead{(6)} & \colhead{(7)} & \colhead{(8)} & \colhead{(9)} & \colhead{(10)} & \colhead{(11)}& \colhead{(12)} }
\startdata
\multicolumn{12}{c}{Shared Envelope} \\
  \hline
MIR1-C1 & -38.5 & 90 & 1.0 & 2.9 & -- & -- & -- & 2.0 & 1.4 & Radiation & \ref{subsec:ice} \\
MIR2-C1 & -38.5 & 43 & 1.0  & 3.1 & -- & -- & 110 & 2.0 & 1.4& Radiation & \ref{subsec:ice}\\
\hline
 \multicolumn{12}{c}{MIR2 Bullets} \\
   \hline
 MIR2-B1 & -89 & 223  & 0.8 & 0.21 & 21 & 7.27& 7.4 & 3.5& 3.2& J-shock & \ref{subsec: bullets}\\
MIR2-B2 & -79 & 180 & 0.75 &  0.32 & 27& 7.60 & 5.3  & 3.0& 2.6 &J-shock& \ref{subsec: bullets}\\
MIR2-B3 & -70 & 221 & 0.65 & 0.12 & 27& 7.60 & 2.0 & 3.0 &2.6&  J-shock& \ref{subsec: bullets}\\
MIR2-B4 & -63 & 325 & 0.7 & 0.14 & 30& 7.73& 1.8 & 4.5& 4.3& J-shock& \ref{subsec: bullets}\\
\hline
\multicolumn{12}{c}{MIR1-MIR2 Immediate Environment (Warm)}\\
\hline
MIR1-W1 & -43  & 180  & 0.6 & 1.4 &  13\tablenotemark{a} & $>$6.52 & $<$289  & 4.0 & 2.0 & Radiation &\ref{subsec:rh}\\
MIR2-W2 & -45.5  & 116 & -- & 3.6 & 13\tablenotemark{a} & $>$6.52 & $<$720 & 1.7 & 0.9 & Radiation & \ref{subsec:rh}\\
\hline
\multicolumn{12}{c}{MIR1-MIR2 Immediate Environment (Hot): Foreground Interpretations} \\
\hline
MIR1-W1$^\prime$ & -46$\sim$-49  & 449 & 0.5 $\&$ 0.2 & 1.7 $\&$ 20.6  &  $\nu$=1-2 & $>$10 & --  &  2.8 $\&$ 3.9 &  2.5 $\&$ 3.7 & -- & \ref{subsec:hot-shock}\\
MIR2-H1 & -37.5 & 756 &  0.4 & 5.8 & $\nu$=1-2 & $>$10 & 0.4  & 4.3  & 4.1& --  & \ref{subsec:hot-shock}\\ 
MIR2-H2 & -52  & 565 & 0.4 & 0.7 & 37& 8.01 & 4.1 & 3.7 & 3.5 & --  & \ref{subsec:hot-shock} \\
\hline
 \multicolumn{12}{c}{MIR1-MIR2 Immediate Environment (Hot): Disk Interpretations} \\
   \hline
MIR1-W1$^\prime$  & -46$\sim$-49  & 709 & --  & 3.7  &  $\nu$=1-2 & $>$10 & -- &  5.0 & 4.8 &  Disk  & \ref{subsec:hot-disk}\\
MIR2-H1 & -37.5 & 662 &  -- & 15.3 & $\nu$=1-2 & $>$10 & -- & 4.3  & 4.1&  Disk  & \ref{subsec:hot-disk}\\ 
MIR2-H2 & -52  & 482 & -- & 0.9 & 37 & 8.01 & -- & 3.7 & 3.5 & Disk & \ref{subsec:hot-disk}\\
[0.5ex] 
\enddata
 \tablecomments{Column (1): Identified components.\\
 Column (5): $N_{\textrm{H}_2}$ = $N_{\textrm{CO}}/(1.6\times10^{-4})$ \citep{cardelli96, sofia97}. \\
 Column (7): $n_\textrm{crit}$ corresponds to the critical density that the highest $J$ level requires to be thermalized. For \co, $n_\textrm{crit}=2\times 10^3 J^3$\percmcu. This is a lower limit when estimating $n$.\\
 Column (8): $d = N_{\textrm{H}_2}/n_\textrm{crit}$; this is an upper limit for estimating the slab thickness. \\
 Column (10): Deconvolved observed velocity dispersion, $\sigma_v = \sqrt{(\sigma_\textrm{obs})^2-(\sigma_\textrm{res})^2}$. $\sigma_\textrm{res}$ = $c/(2\sqrt{2\textrm{ln}2}R)$.}
\tablenotetext{a}{$J_\textrm{max}$ identified in \thco; this is a very underestimated value, because \co\ data at this velocity is saturated and not usable. For \thco, we use $n_\textrm{crit}=1.5\times 10^3 J^3$\percmcu.}
\end{deluxetable*} 

The picture that emerges from the M-band spectroscopic study is of a shared foreground envelope, several high-velocity clumps, {and} a few warm or/and hot components in the immediate environment of the binary (Table~\ref{table:sum}). Identification of the origin of the absorption components requires more extensive analysis. In the subsequent subsections, we will place the observed CO absorption components in the framework of the known structures in the \wt\ star-forming region.

\begin{figure*}[!t]
    \centering
    \includegraphics[width=\linewidth]{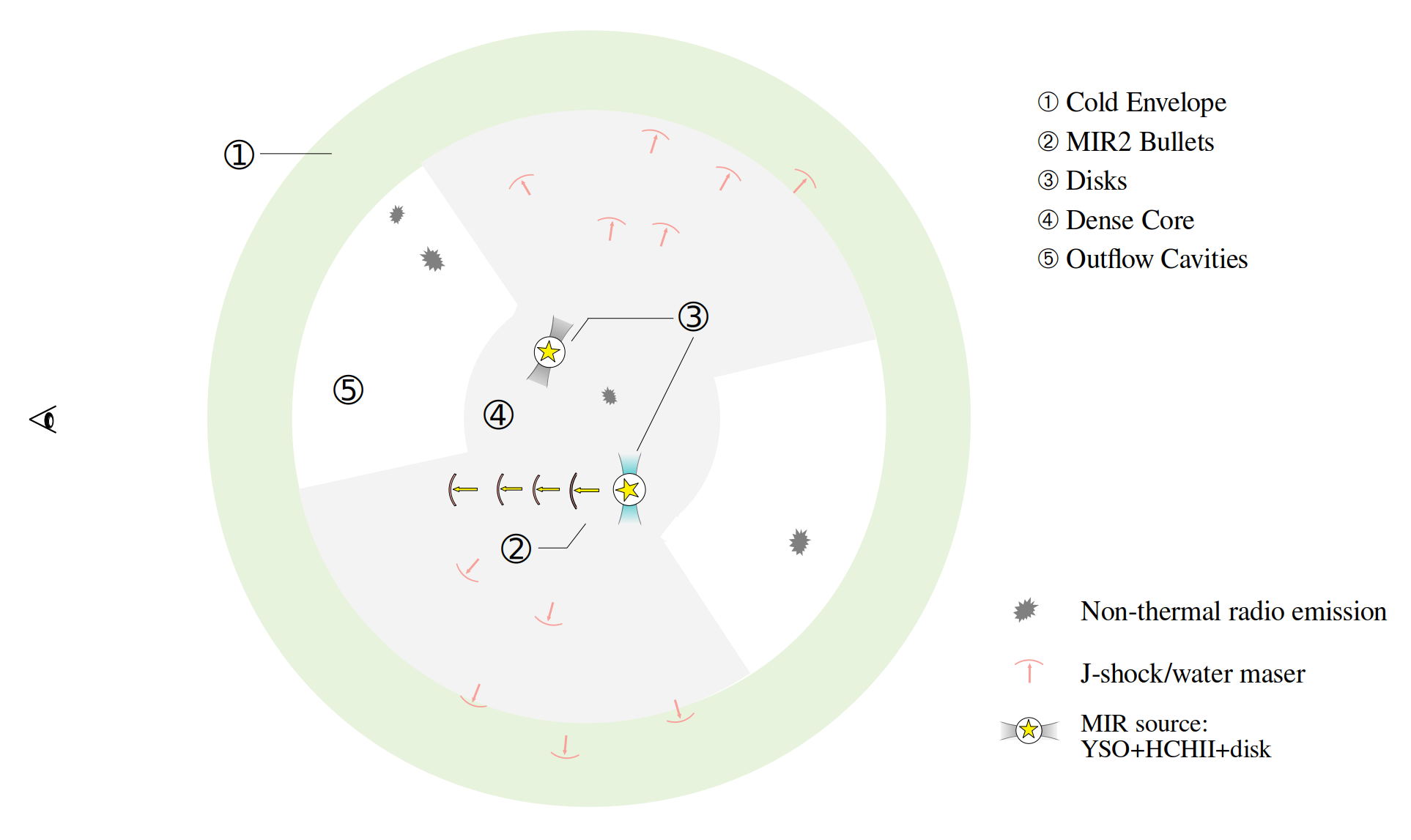}
    \caption{Schematic view of the potential environment of \wt. {Important structures from past studies including the envelope, the conical region cleared by the outflows, the non-thermal continuum sources, and water masers are presented. We locate the origins of new structures we identify in CO absorption such as high-velocity bullets and disks against this backdrop.}}
    \label{fig:model2}
\end{figure*}

\subsection{Known Structures of \wt}\label{subsec: wt-past}

\wt\ is a very active region of massive star formation and the binary is oriented along the northeast-to-southwest direction. The binary is enclosed by a 10$^4$~AU hot core detected in CS \citep{vdt00} and a rotating toroid or envelope detected in SO$_2$ \citep{rodon08, wang12, wang13} of a similar size. Outflows were observed at different scales: a bipolar outflow in CO(2--1) along the northeast-to-southwest direction was observed by JCMT \citep[$>$10$^5$~AU;][]{mitchell91} ranging from $v_{\rm LSR}$ of $-20$ to $-60$~\kms, and two outflows along the line of sight were detected in SiO(5--4) \citep[0.39\asec$\times$0.34\asec\ beam at 1.4~mm,][]{rodon08} by PdBI from $v_{\rm LSR}$ of ${-30}$ to ${-50}$~\kms. {A cavity in front of the binary cleared by the outflows was suggested to exist due to the low estimated foreground extinction \citep{vdt05}.} Along the northeast-to-southwest direction, a few fast-moving, compact non-thermal radio continuum sources were found. As these jet-lobes indicate jet-disk systems ($\sim$10$^3$~AU), MIR1 and MIR2 with thermal radio emission are disk candidates \citep{wilson03, purser21}. Moreover, hundreds of water maser spots are widely spread in the same region \citep{menten90, imai00}, suggesting active clumps-surrounding gas interactions in the nearby region to the binary.

We present in Figure~\ref{fig:model2} the envelope, the conical region cleared by the outflows, the non-thermal continuum sources, and water masers to illustrate the important structures of \wt. It is against this backdrop that we have to identify the origin of the different absorption components observed at mid-IR wavelengths in MIR1 and MIR2.

\subsection{The Foreground Envelope: Gas and Ice}\label{subsec:ice}

\begin{figure*}
    \centering
    \includegraphics[width=\linewidth]{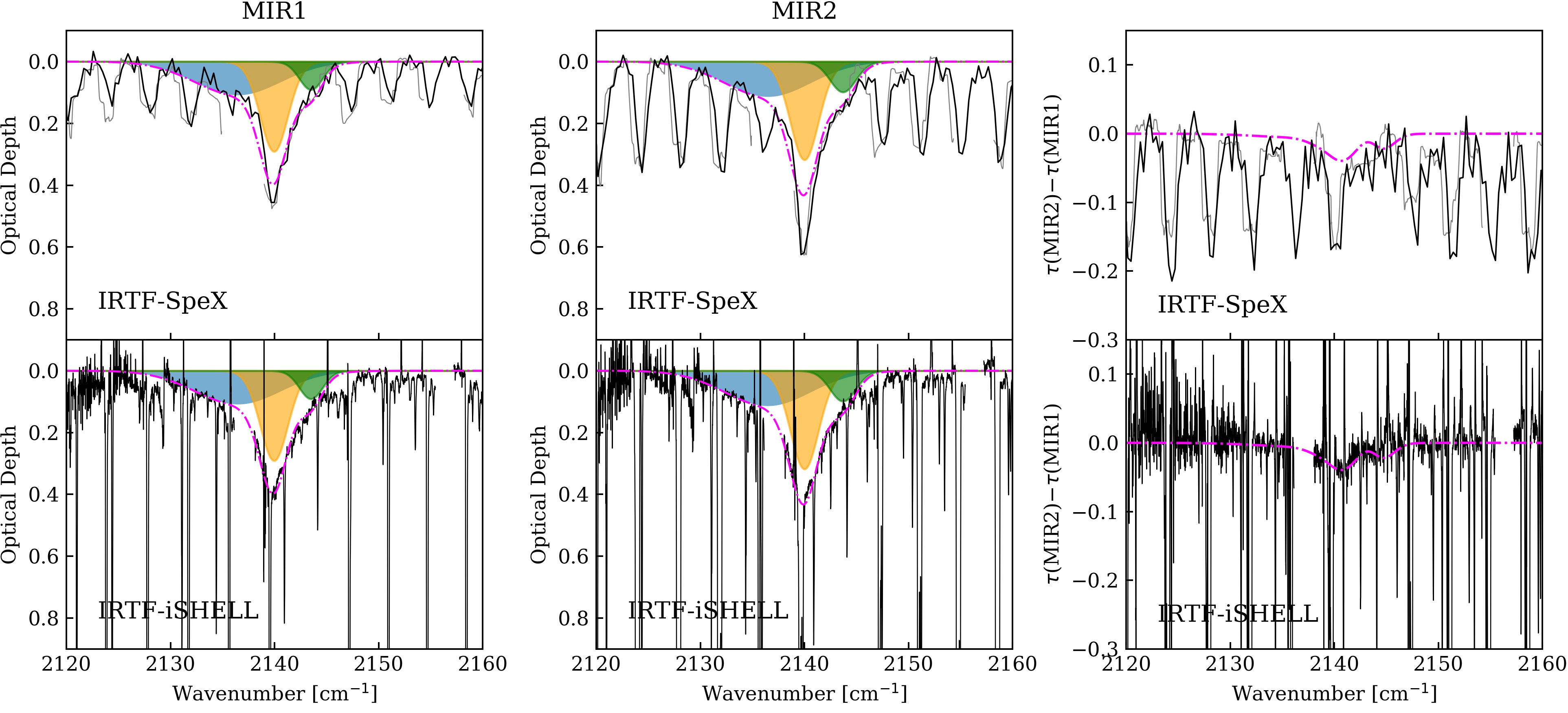}
    \caption{\textcolor{black}{IRTF-SpeX (\textit{upper panel}; $R$=1200) and IRTF-iSHELL spectra (\textit{bottom panel}; $R$=88,100) towards MIR1 and MIR2 at 4.67~\um\ showing the CO ice band. IRTF-iSHELL spectra convolved to the resolution of 1200 (\textit{grey}) are overlaid on SpeX spectra. The ice absorption profiles are fitted by three Gaussians over the iSHELL spectra: non-polar CO (\textit{blue}, 2136.5~cm$^{-1}$), polar CO (\textit{yellow}, 2139.9~cm$^{-1}$), non-polar CO (\textit{green}, CO$_2$/CO$>$1, 2143.7~cm$^{-1}$). Gaussian fittings on iSHELL spectra are overlapped on SpeX spectra, as gaseous CO absorption lines there contaminate the ice absorption. The bottom right panel shows that the difference of the optical depth of features between MIR1 and MIR2 is smaller than 5\%.}}
    \label{fig:ice}
\end{figure*}

\begin{figure*}[!t]
    \centering
    \includegraphics[width=\linewidth]{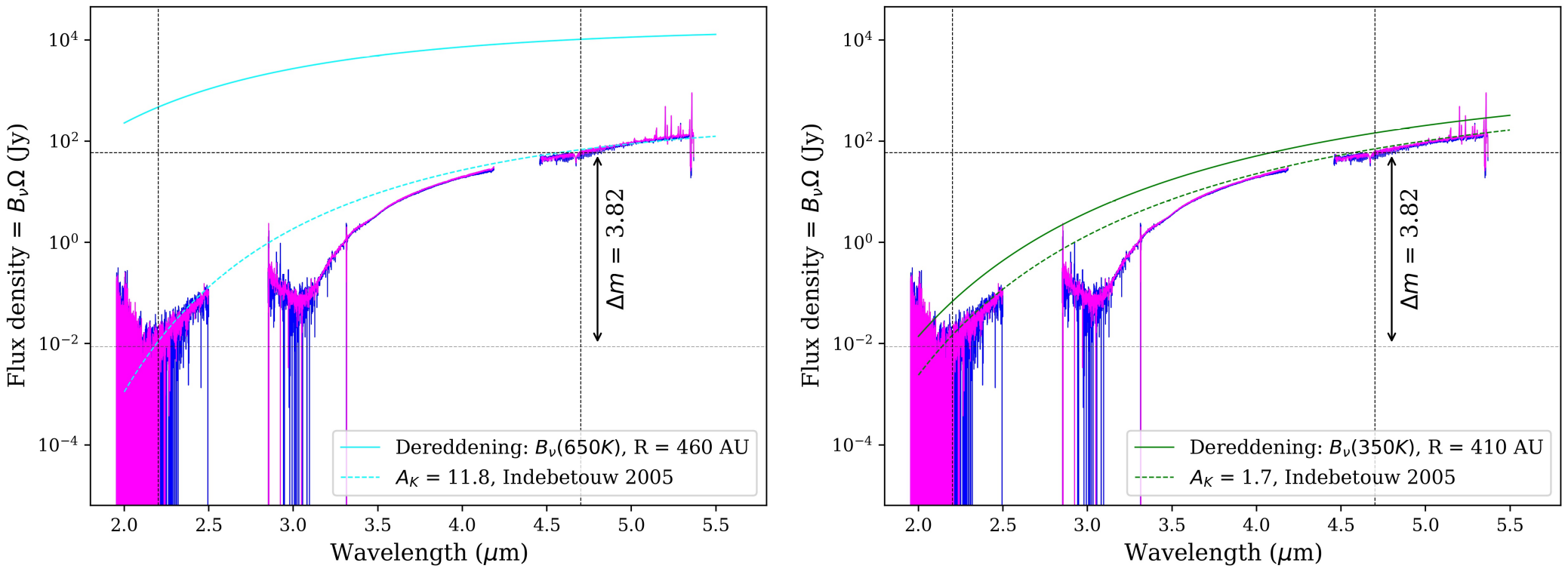}
    \caption{IRTF-SpeX spectra from 2--5~\um\ of MIR1 (\textit{magenta}) and MIR2 (\textit{blue}) compared to the flux density from two disks emitting in black body. Both the black body model (\textit{solid}) and the reddened spectra (\textit{dashed}) following the extinction law in \citet{inte05} are presented. The extinction on the left panel is consistent with the 9.7~\um\ optical depth ($A_K = 11.8$), and that on the right panel is consistent with the column ($A_K = 1.7$) derived from the gaseous CO absorption adopting a CO abundance of 1.6$\times10^{-4}$.}
    \label{fig:spex-con}
\end{figure*}

Because MIR1-C1 and MIR2-C1 have almost the same column density ($N_{^{13} \rm{CO}}\simeq 7.2\times10^{16}$~\percmsq; Table~\ref{table:cor-slab}) and are at the same velocity, it is reasonable to designate them in the envelope surrounding the binary. The derived temperatures are cool ($\sim$40--90~K), suggesting that the envelope layer is rather far away from the protostars.

The total H column density of the envelope can be derived to be $N_H=2\times10^{23}$~\percmsq~($A_V = 108$) from the observed 9.7~\um\ silicate optical depth \citep[$\tau_{9.7~\mu m} \simeq 5.8$;][]{gibb04}, the $A_V/\tau_{\rm sil}$ = 18.6, and $N_H/A_V =1.9\times10^{21}$~\percmsq~\citep{roche84, bohlin78}. Therefore, adopting a $^{12}$C/$^{13}$C elemental abundance ratio of 65~\citep{milam05}, this H column density ($A_V$ = 108) implies an abundance of gaseous CO in the envelope of $2\times10^{-5}$. With a gas phase C abundance of $1.6\times10^{-4}$ \citep{cardelli96, sofia97}, gaseous CO is not the main reservoir of carbon along these sight-lines. 

An independent view on the gas and dust columns towards W3 IRS5 is given by the IRTF/SpeX 2--5~\um\ micron spectra. We find that the foreground column density is consistent with the 9.7~\um\ optical depth if we consider that the near and mid-IR continuum originates from the
blackbody emission of the disks. Figure~\ref{fig:spex-con} shows that the magnitude difference $\Delta m$ between the $K$ and $M$-bands is 3.82. Adopting the extinction curve from \citet{inte05} and $A_K = 11.8$ ($A_V = 108$) derived from the 9.7~\um\ silicate band, the spectrum is consistent with a disk in a radius of 460~AU emitting at 650~K, a typical temperature we have derived for the hot CO components (see Table~\ref{table:sum} and \S~\ref{subsec:hot-shock}). On the contrary, if we use the foreground extinction that we measured from gaseous CO~($A_K = 1.7$, or equivalently, $\sim3\times10^{22}$~\percmsq; Table~\ref{table:sum}) adopting the canonical \co\ abundance of $1.6\times10^{-4}$ (e.g., all the gas phase C in CO), this leads to a blackbody temperature of 350~K with a radius of 410~AU. However, we consider that the latter disk model is less likely because hot components above 600~K (e.g. MIR2-H1) will have emission rather than absorption lines against the 350~K disk (see Appendix A in Barr et al. 2022, \textit{submitted}). Hence, this analysis of the near-infrared spectral energy distribution also implies that the measured CO column density of the cool foreground gas is only a good measure of the total hydrogen column density if we adopt a low ($2\times10^{-5}$) abundance for \co. We suggest that mid-IR interferometry observations may be able to distinguish between the different models at much lower temperature and much smaller foreground extinction. 

The iSHELL/IRTF provides a direct handle on the solid CO ice along the same sight-lines {(see Figure~\ref{fig:ice})}. The CO ice absorption profiles towards MIR1 and MIR2 are almost identical, supporting our conclusion that there is a shared cool envelope in front of the binary. Adopting a band strength $A = 1.1\times10^{-17}$ \citep{pon03}, $N_\textrm{CO, \textrm{ice}}$ = 2.1$\times$10$^{17}$~cm$^{-2}$. Specifically, non-polar CO (centered at 2136.5~cm$^{-1}$) and polar CO (centered at 2139.9~cm$^{-1}$) have comparable column density, suggesting that half of the solid CO ice is in H$_2$O- or CH$_3$OH-rich ice. Moreover, taking the column of CO$_2$ ice, 7.1$\times10^{17}$~cm$^{-2}$ \citep{gibb04}, into account of the carbon inventory, carbon in solid phase is 19.6\% of H$_2$O ice, and 19.8\% of gaseous CO. The column of H$_2$O ice is measured through the 3~$\mu$m absorption feature and is $5.1\times10^{18}$~cm$^{-2}$ \citep{gibb04}. Hence, the identified carbon-bearing ice species cannot account for the missing C in the envelope. We may speculate that prolonged UV photolysis has converted the carbon containing ice compounds into an organic residue \citep{bs95, bs97, vino13}.

Taking into account that dust and gas will be well coupled thermally, we note that the derived temperature of the cold gaseous component (40--90 K) is well above the sublimation temperature of pure CO. Likewise, much of the CO trapped as a trace species in H$_2$O-ice will sublimate at a temperature of $T_d\simeq$ 50 K. We surmise that the polar CO-ice component coexists with the gaseous CO MIR1- and MIR2-C1 components in the warmer parts of the envelope while the apolar component likely resides further away in a region where dust temperatures are below 20 K
\citep[see Figure~7.9 in ][]{tielens21}.

\subsection{The Foreground Bullets of MIR2}\label{subsec: bullets}

Although sub-millimeter observations have seen multiple molecular outflows in W3 IRS5, none of them are direct counterparts of MIR2-B1 to B4, which are at $\sim$20--50 \kms\ relative to $v_{sys}$ of $-38.5$~\kms\ (see Section~\ref{subsec: wt-past}). The range of the radial expansion velocity of water masers observed at 22~GHz \citep[up to 60~\kms;][]{imai00}, on the other hand, is comparable with that of the MIR2 ``bullets". As maser action is limited to regions of long velocity coherence, the bullets could be related to the water masers if they are directed more toward us and their maser action directed away from us. In the remainder of this section, we will examine this possibility. As maser emission originates from either J-type or C-type shocks driven by protostellar outflow \citep{hem13}, we will consider the implications of each of these possibilities separately.

\subsubsection{A J-Shock Origin}

In fast {J-shocks}, the gas is instantaneously heated to an extremely high temperature up to 10$^5$~K that completely dissociates molecules and partially destroys dust behind the shock. As the material cools down, H$_2$ reforms on the surviving dust and is collisionally de-excited. The H$_2$ re-forming stage provides a heating source and maintains the gas in a temperature plateau of $\sim$300--400~K \citep{hem13}. This warm gas is very conducive to molecule formation and the H$_2$O maser emission and CO absorption could originate in this temperature plateau region.

We {can} compare the physical properties of the absorbing bullets with what the J-shock model would predict. First, in the J-shock model for the H$_2$O maser emission, the CO column density of the heated plateau is as large as 3$\times10^{17}$ to $3\times10^{18}$~\percmsq~\citep{neufeld94}, and is consistent with the derived column densities of MIR2-B1 to B4 in Table~\ref{table:sum}. Second, the post-shock plateau density can be as high as 10$^8$--10$^9$~\percmcu~\citep{hem13}, and is also compatible with that of the bullets. As we have observed energy levels in LTE up to $J$ = 30, the corresponding critical density is $\sim 3 \times 10^3 \,J^3 \sim 8.1 \times 10^7$~\percmcu. This is a lower limit. In short, the temperature, the column density, and the particle density are in the favored parameter space for the masing regions produced by J-shocks.

\subsubsection{A C-Shock Origin}\label{bullet-c-shock}

In contrast to J-shocks, H$_2$ is not dissociated in C-shocks, and its relatively constant temperature plateau is kept by the frictional heating between ions and neutrals. The temperature of the shocked gas is typically much higher \citep[$>$1000~K;][]{kaufman96} than observed for the MIR2-B1 to B4 CO absorption components. In the masing plateau, the warm hydrogen column is $\sim$10$^{21}$~\percmsq~\citep[$1.6\times 10^{17}$~\percmsq\ for CO;][]{kaufman96}. Similar to J-shocks, densities of 10$^8-10^9$ \percmcu\ are required for H$_2$O maser action in C-shocks \citep{hem13}. 

Therefore, interpreting the warm gas in the B components in MIR2 as C shocks faces some issues. First, the observed temperature under 400 K is rather low and would restrict the shocks to a velocity less than 10 km/s \citep{kaufman96}. Second, the observed column densities are a factor 1--3 larger than C-shocks can produce. This is irrespective of whether these bullets are truly water maser counterparts.

\subsubsection{Linking the Bullets to Water Masers}

If MIR2-B1 to B4 indeed originate from the post-J-shock gas, the physical properties we obtain along the line of sight direction are complementing what water masers convey on the sky plane. The CO absorption lines and the water masers are depicting different perspectives of the geometry of the post-shock gas. Water masers are beamed emissions that require enough coherence path length in our direction. While masers formed in a compressed shell of post-shock gas swept up by outflows, observable masers are typically viewed from an ``edge-on" direction that is perpendicular to the motion of the shock. Therefore, while the brightest masers have the lowest line-of-sight velocities, CO components detected in our absorption spectra complement information along the line of sight, and the maser emissions are the weakest.

The water masers have a smaller velocity width and size than the B components: in Table~\ref{table:sum}, the velocity widths (2.6--4.3~\kms) and the thickness (2--10~AU) of the absorbing bullets are larger than but not incompatible with the average velocity width, from 0.8--1.6~\kms\ in 1997 and 1998, and the average size of 1~AU of 22~GHz water masers observed in the region surrounding MIR1 and MIR2 in W3~IRS5 \citep{imai02}. We may expect that the water masers have a smaller velocity width because of the required velocity coherence. We are observing them perpendicular to the propagation direction, while the bullets are coming more toward the observer. Besides, the estimated thickness of the bullet is an upper limit derived simply by $N_{\textrm{H}_2}/n_{\textrm{crit}}$. As a reference, the thickness of the masing region predicted by the shock model \citep{hem13} is $\sim 10^{14}$~cm (6.7~AU). \citet{hem13} has proved that, although the value of the maser thickness predicted by the J-shock model depends on exact shock properties, strong water maser emission is a robust phenomenon that can be generated from a wide range of physical conditions without a fine-tuning of parameters.

We have to consider the likelihood that the four bullets are intercepting the narrow pencil beam set up by the background IR source. For the hundreds of water masers, \citet{imai02} found that the two-point spatial correlation function among 905 maser spots can be fitted by a power-law. With an index of $-2$ in a linear-scale range of 1.1--540~AU, this indicates that in this scale range the features are clustered and have a ``fractal" distribution. Considering that adding four more radially identified data points (MIR2-B1 to B4) over the  0.375$^{''}\times1.2^{''}$ continuum from our observations does not have a significant influence over the correlation function, the correspondent $\sim$10 spots per square arcsecond corresponds to a spatial separation of $\sim$0.2\asec ($\sim$450~AU), which fits with the maser geometry \citep{imai00}. However, we note that those seemingly nice fits do not answer why the spatial distribution of the masers is clustered. As for the direction along the line of sight, \citet{imai02} also found a power-law on the velocity correlation function. The measured difference in Doppler velocity and the spatial separation was fit with an index of 0.29$\pm$0.03 and was putatively linked to Kolmogorov-type turbulence in the interiors of the masers. It was suggested that small-scale turbulence was left in the subsonic part of the post-shock region \citep{gwinn94}. MIR2-B1 to B4 fit into this power-law correlation, consistent with a post-shock origin.

\subsection{The Immediate Regions of MIR1 and MIR2}\label{intermediate}

All absorbing components in our mid-infrared spectra other than MIR2-B1 to B4 are located in the range $\sim-38$ to $-60$~\kms. Although our analyses of the isotope lines have shown that the components within this range are different, saturated \co\ low-J lines still share a fortuitous similar line profile, with its red edge contributed by MIR1-C1 and MIR2-C1, the surrounding envelope at $v_\textrm{sys}$, and its blue edge contributed by MIR1-W1/W1$^\prime$ and MIR2-H2. Such a profile shared by MIR1 and MIR2 indicates the underlying correlation of the two sources, and we investigate how components within $\sim-38$ to $-60$~\kms\ constitute the immediate regions of MIR1 and MIR2. As the observations are consistent with either absorption arising in foreground clumps or in the disk, we will consider these in turn.

\subsubsection{Radiative Heating}\label{subsec:rh}

Assuming the gas and the dust are thermally coupled, we use equation 5.44 and equation 5.43 in \citet{tielens05} to estimate the distance of the gas to the protostars if the gas is radiatively heated:
\begin{equation}
    {T_d} \simeq 53 \left( \frac{0.1~ \mu m}{a}\right)^{0.2}\left( \frac{G_0}{10^4}\right)^{0.2},
\end{equation}
and
\begin{equation}
    G_0=2.1\times 10^4 \left( \frac{L_*}{10^4 L_\odot}\right)\left( \frac{0.1~\textrm{pc}}{d}\right)^{2},
\end{equation}
in which $G_0$ is the radiation field in terms of Habing field, $a$ is the grain size, $L_*$ is the stellar luminosity, and $d$ is the distance. Taking 0.1~\um\ as a typical size for interstellar grains\footnote[1]{{We recognize that grains may have grown to $\sim$0.3--0.5~\um\ in dense clouds due to coagulation~\citep{ormel11}, but that will have a very little effect on the mid-IR absorption compared to the far-IR emission. We estimated it will change the temperature by 20--30\%.}}, and that MIR1 and MIR2 having a similar $L_*$ of 4$\times10^4 L_\odot$~\citep{vdt05}, we arrive at a $d$ of 280, 140~AU for the 450~K (MIR1-W1$^\prime$) and $\sim$600~K (MIR2-H1, MIR2-H2) components, and $d$ of at least 2000~AU for the cooler components (MIR1-W1, MIR2-W2, $<200$~K). 

However, locating the hot components at such a small distance to the protostars is in conflict with the similarity of the 1991 and 2018 CO absorption line profiles (see \S~\ref{subsec:slab-h1h2}). Considering that MIR1-W1$^\prime$ and MIR2-H2 have a constant relative velocity of 10~\kms\ and 15~\kms, the two components moved outwards along the line of sight for 60 and 100~AU in the past 30 yrs. As the moving distances are quite large, radiative heating cannot keep MIR1-W1$^\prime$ and MIR2-H2 at the observed high temperature. As for MIR2-H1 which is at $v_\textrm{sys}$, it would have to stay static at a distance of only 140~AU for 30 yrs and yet be close to a massive forming protostar. It could be part of a disk associated with the protostar.

\subsubsection{A J-shock/C-shock Origin}\label{subsec:hot-shock}

Alternatively, a shock origin for the hot components is attractive, as the vibrationally excited lines observed towards MIR1-W1$^\prime$ and MIR2-H1 indicate a very high density, $\sim10^{10}$~\percmcu\ for thermalization at the $\nu$=1 level. However, because J-shocks cannot heat the masing gas to temperatures greater than about 400~K~\citep{hem13}, and the column density is far too large to be consistent with C-shocks (see \S~\ref{bullet-c-shock}), neither a J-shock nor a C-shock model fits.

\subsubsection{A Disk Origin}\label{subsec:hot-disk}

\begin{deluxetable*}{c c c c c cc }[!t]
\tablecolumns{6}
\tabletypesize{\scriptsize}
\tablecaption{Keplerian Parameters of Hot Blobs on the Disk \label{table:kep}}
\tablehead{\colhead{Component} & \colhead{$v_{lsr}$(MIR1)}& \colhead{$v_{lsr}$(MIR2)} &\colhead{$v - v_{lsr}$}   & \colhead{$d$}  & \colhead{$P$} & \colhead{Note}\\
\colhead{} & \colhead{(\kms)}   & \colhead{(\kms)}   & \colhead{(\kms)}   & \colhead{(AU)}  & \colhead{(yrs)} & \colhead{}}
\startdata
MIR1-W1$^\prime$ & -38 & -- & $>$10 & $<$180& $<$540& Blob on an inclined disk \\
MIR1-W1$^\prime$ & -46 & -- & 0 & -- & -- &  Annular structure \\
MIR2-H1 & -- & -38 & 0  &--& -- & Blob on a face-on disk\\
MIR2-H2 & -38 & -38 &$>$15  & $<$80 & $<$180 &  Blob on an inclined disk \\
\enddata 
\end{deluxetable*} 

In Section~\ref{subsec:cog-star}, we have illustrated that the curve of growth analysis on a disk model can reconcile the temperatures measured from the observed CO isotope spectra of MIR1-W$^\prime$, MIR2-H1, and MIR2-H2 in a 1$\sigma$ level. Other than being a feasible model, such a disk origin of hot gas has been proposed in other massive protostellar systems. Take AFGL~2591 and AFGL~2136 as examples, in \citet{barr20}, absorption features against the mid-IR continuum were detected in CO, CS, HCN, C$_2$H$_2$, and NH$_3$, and all have a temperature of $\sim$600~K. The disk origin was motivated by the abundance difference on both HCN and C$_2$H$_2$ at 7~\um\ and 13~\um; e.g., The abundance derived from HCN as well as C$_2$H$_2$ lines at 13~\um\ is about an order of magnitude smaller than those derived from lines at 7~\um\, even though the lines originate from the same ground state. This was attributed to a dilution effect at 13~\um\ as the outer parts of the disk radiate predominantly at 13~\um\ and these outer layers have lower abundances of these species \citep{barr20}. Mid-IR interferometry has shown that the IR emission originates from a structure with a size of $\sim$100--200 AU for both AFGL~2136 {\citep{monnier09, dewit11, frost21}} and AFGL~2591 {\citep{monnier09, olguin20}} and likely this is a disk. This scenario is also supported by ALMA \citep[AFGL2136;][]{maud19} and NOEMA {\citep[AFGL2591;][]{suri21}} observations in which Keplerian disks are revealed. Clumpy substructures that may be associated with the absorbing components were resolved on the disk of AFGL~2136 in the 1.3~mm continuum \citep{maud19}, supporting the model that a cooler component is absorbing against the continuum from the disk.

We hereby attribute the absorption to blobs in a disk in accordance with studies of other massive protostars. However, given the unknown inclination and the unknown systematic velocity of the disk, the location of the blobs on the disk is difficult to pinpoint. We present the Keplerian parameters of the three hot components in Table~\ref{table:kep} to illustrate the difficulty in interpreting the kinematics, and specifically note that we measure a radial velocity and the blobs could be much further away if the disks are not in the plane of the sky and radial velocities contain little information on Keplerian motion. We emphasize that the velocity difference between the blobs in MIR1 and MIR2 is the interplay of the orbital motion of the blobs in these disks and the difference in space motion between the two protostars where we note that the orbital motion of a double star system (each has 20~$M_\odot$) at 1000 AU is $\sim$ 3~km/s. Therefore, disks in MIR1 and MIR2 need to be spatially and spectrally resolved to a fully understand of the structures in this region.

We stress in the end that, while such a model is feasible to interpret the observed absorption profiles, we still lack definite evidence to link the absorbing gas in the mid-IR with the disks in \wt. We recall that in Section~\ref{subsec:ice}, for both MIR1 and MIR2, we fit the 2--5~\um\ spectrum with a 650~K disk in a radius of 360~AU. We acknowledge that this is an oversimplified model, and the dust composition, the extinction correction, or the actual disk geometry may influence the fitting result. This radius is compatible with the disk radii (500--2000~AU) that \citet{frost21b} find for some massive young stellar objects using multi-scale and multi-wavelength analysis. However, we recognize that the derived radius is slightly larger than the size of MIR1 and/or MIR2 of 350–500 AU at 4--10~\um\ \citep{vdt05}. {This value is also quite large compared to the $\sim$100~AU size measured by mid-IR interferometry for other massive protostellar systems \citep{monnier09, beltran16}, although \citet{frost21b} discussed that the mid-IR emission is mostly dominated by emission from the inner rim of the disk, therefore may not constitute the size of the whole disk.} We suggest that complementary observations in mid-IR and millimeter interferometry will help to disentangle the issues above. 

\subsection{{Comparison with Hot Core Tracers}}

{The hot core at $v_\textrm{sys}$ of $-38$~\kms\ in the \wt\ system revealed by sub-millimeter molecular lines is a spatially (10$^3$--$10^4$~AU) and spectrally ($\sigma_v\sim$5~\kms) extended structure \citep{vdt00, wang13}. As a comparison, the absorbing components in the mid-IR are observed in ``pencil" beams (sub-arcsec scale; or a few hundreds AU). Since we have decomposed the different CO absorbing components by their velocities and temperatures (Table~\ref{table:sum}), it is of interest to compare the molecular components in emission and in absorption. We note that the post-shock bullets do not leave any signatures on the hot core tracers, possibly because their beam-averaged column densities in the large sub-millimeter beams are very small.}

{The sub-millimeter CO observations reveal emission at $-38$~\kms\ with a \co\ column density derived from \cso\ observations of $3.7\times10^{19}$~\percmsq. This column density is much higher than that of the cold envelope, $-38$~\kms\ components (MIR1-C1/MIR2-C1 and MIR2-H1) probed in the mid-IR ($N_{^{12} \rm{CO}}$ = $4.7\times10^{18}$~\percmsq, Table~\ref{table:cor-slab}, $^{12}$C/$^{13}$C = 65). This may well be because the sub-millimeter includes emission from the core (Figure~\ref{fig:model2}) which is not traversed by the mid-IR pencil beam.}

{Other sub-millimeter molecular tracers such as SO, HCN, and CS rotational transitions also reveal the hot core at a rather extended scale of $\sim$3000--5000~AU \citep[beam size of 1.1\asec$\times$0.8\asec,][]{wang13}. While the exact measurements of the column densities are lacking as the lines are heavily filtered out at $v_\textrm{sys}$, these tracers are all in the velocity range ($-30$ to $-50$~\kms) characteristic of the molecular core. The sub-millimeter continuum dust emission provides a beam averaged, H$_2$ column density of the core of $1.5\times10^{23}$~\percmsq~\citep{wang13}. Coincidentally, this is similar to the H$_2$ column density derived from the envelope derived from the pencil beam observations of the strength of the 9.7~\um\ silicate feature ($2\times10^{23}$~\percmsq, Section~\ref{subsec:ice}). Therefore, similar to the discussion of CO emission lines above, neither the SO, HCN, and CS rotational lines nor the dust continuum traces the envelope components MIR1-C1/MIR2-C1 and MIR2-H1 probed in the mid-IR but rather entire core region.}

{The SOFIA HyGal survey \citep{jacob22} provides constraints on hydride molecules (such as CH) and atomic constituents (C$^+$ and O) against the far-IR/sub-millimeter continuum as well with beam sizes of from 6--14\asec. CH, C$^+$, and O are all in the velocity range of the envelope. Adopting a CH abundance of 3.5$\times10^{-8}$, appropriate for diffuse clouds~\citep{sheffer08}, \citet{jacob22} infer an H$_2$ column density of $2.7\times10^{21}$~\percmsq. Even if we adopt an abundance of $10^{-8}$, typical for dark cloud cores \citep{loison14}, the inferred H$_2$ column density is only $10^{22}$~\percmsq. This is small compared to either the pencil beam column density derived for the envelope from the 9.7~\um\ silicate feature or the average column density of the core derived from the sub-millimeter dust. Perhaps, much of the carbon has frozen out in the envelope and/or core in the ice mantles. The CH observed by HyGal may then mainly trace the surface layers of the cloud. It is reasonable to assume that the [CII] 1.9 THz fine-structure line traces the photo-dissociated surfaces of the molecular cloud. Taking a fractional gas-phase carbon abundance  of 1.6$\times10^{-4}$~\citep{sofia97}, this corresponds to a column of hydrogen of 3.7$\times10^{21}$~\percmsq, a typical value for a PDR surface \citep{tielens85}. The column density of O measured at 63~\um~(2.2$\times10^{18}$~\percmsq) is rather comparable to the amount of oxygen in water ice (5$\times10^{18}$~\percmsq) measured at 3~\um, and a large fraction of the elemental oxygen is locked up in water ice in the envelope.}



\section{Summary}

In this paper, we report the results from a high resolution ($R$=88,100) mid-infrared spectroscopy study of \wt\ at 4.7~\um, in which hundreds of absorption lines of \co\ and its isotopologues, including \thco, \ceio, \cso\ were resolved. The main results are summarized as follows:

\begin{itemize}
    \item Different spectroscopic properties of MIR1 and MIR2 are spatially resolved for the first time, and the high-velocity components between -60 to \mbox{-90}~\kms\ are attributed exclusively to MIR2.
    
    \item MIR1 and MIR2 share very similar saturated \co\ line profiles between -38 to -60~\kms\ in low-J lines, but we decomposed and identified components from the blended profiles with very different physical properties.
    
    \item For components identified with Gaussian fittings, their physical conditions derived from the rotation diagram analyses show that optical thin assumptions fail. The derived column density ratios are much lower than the expected CO isotope ratios, indicating that optical depth effects have affected the rotation diagram analyses.
    
    \item To reconcile the physical properties derived from different isotopes from the same velocity component, we have analyzed the data using a curve of growth approach. In this, we consider two scenarios:  (1) absorption by foreground blobs that partially cover the background continuum source. (2) Absorption in the photosphere of a circumstellar disk that has a decreasing temperature gradient in the vertical direction.

    \item We applied the slab model on all the components and constrained the corresponding covering factor and Doppler width. We found that this slab model fits nicely to most of the components other than two very hot ones with large column densities (MIR1-W1$^\prime$ and MIR-H2). 
    
    \item We applied the stellar atmosphere model to all the hot components ($>$400~K) and were able to reconcile all the related molecular lines to a single curve. This procedure provides abundance ratios relative to the mid-IR continuum opacity of the dust.
    
    \item We assign the identified components to the immediate environment of \wt, including the shared envelope, the foreground clumps produced by either J- or C-shocks, and the disk. Direct radiation can be a heating mechanism for some components.

    \item MIR1-C1 and MIR2-C1 originate from a shared cool envelope in front of the binary. However, the rather low abundance of gaseous CO suggests that gaseous CO is not the main reservoir of carbon in the envelope. The identified carbon-bearing ice species cannot account for the missing C in the envelope.
    
    \item MIR2-B1 to B4 (``bullets") are possibly J-shock-compressed regions akin to the regions that produce the water maser emission. Our observations in CO lines likely complement the constraints on the physical conditions of water masers from a different geometry perspective. As bright water maser spots are usually beaming in a direction that is perpendicular to their motions, CO absorption lines reveal their properties along the line of sight when water masers have the weakest brightness.  
    
    \item The modeled spectra of MIR2-H1 from both modifications do not match the observed spectra of MIR2 in its red wing at $v_\textrm{sys}>-38$~\kms. However, we found that the residual between the model and the measurement matches the potential emission component reported by \citet{mitchell91} at $-35$~\kms\ in velocity position and intensity. If the residual represents a real emission component, this is a P Cygni profile indicative of an outflow on a scale of $\sim$1000 AU.

    \item Our curve of analyses favor the hot components (400--700~K) located at the two circumstellar disks. However, given the unknown inclination and the unknown systematic velocity of the disk, the location of the blobs on the disk is difficult to pinpoint. Spatially and spectrally resolving the disks in MIR1 and MIR2 will help fully understand the structures in this region.

\end{itemize}

We acknowledge the anonymous referee for providing helpful suggestions to improve the quality of this paper. Support for the EXES Survey of the Molecular Inventor of Hot Cores (SOFIA $\#$08-0136) at the University of Maryland was provided by NASA (NNA17BF53C) Cycle Eight GO Proposal for the Stratospheric Observatory for Infrared Astronomy (SOFIA) project issued by USRA.

\software{Astropy \citep{astropy13, astropy18}, NumPy \citep{harris20}, SciPy \citep{virtanen20}.}

\appendix

\section{Additional Figures}

\begin{figure*}[!h]
\centering
    {\includegraphics[width=0.9\linewidth]{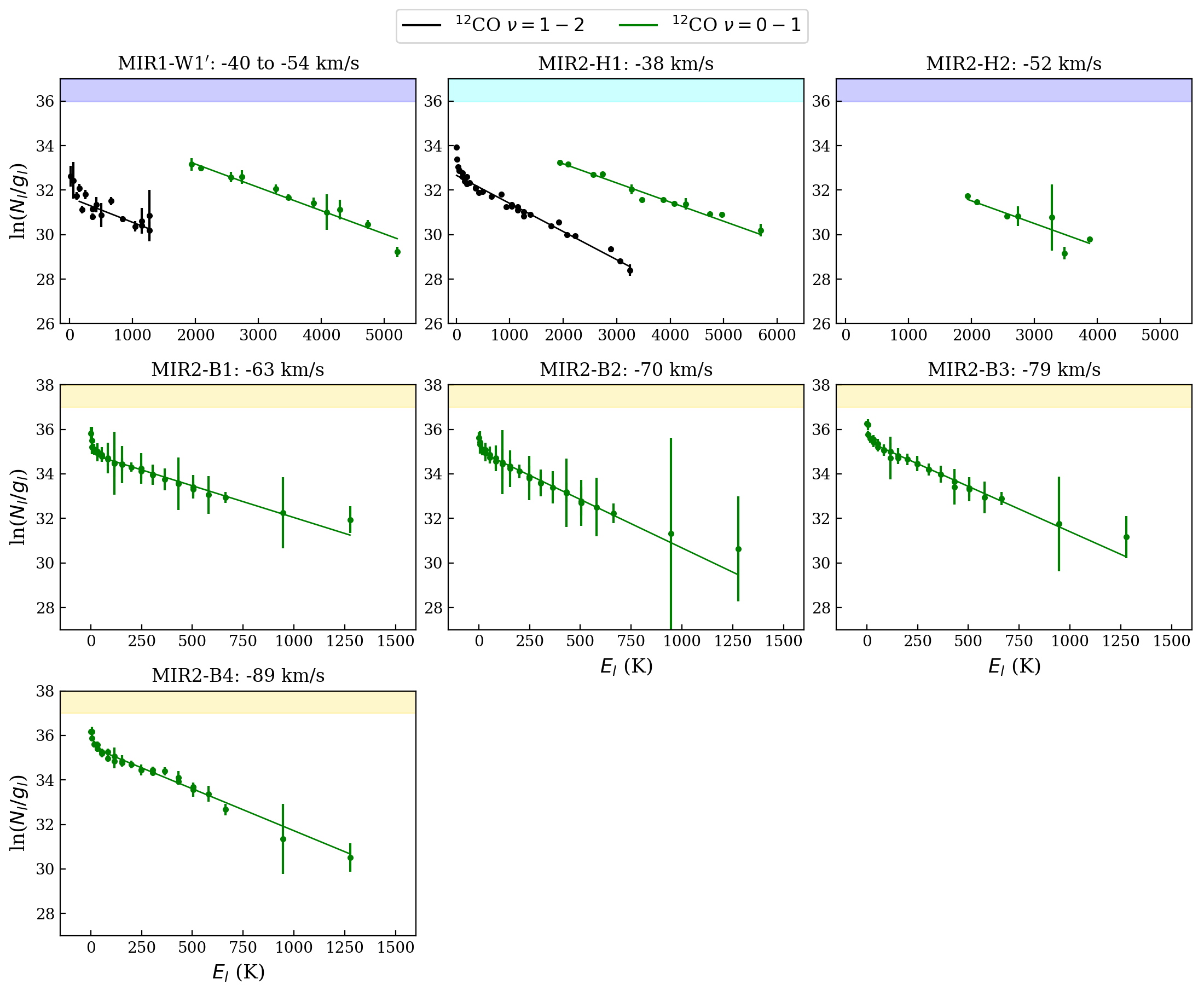}  }
    {\caption{Supplementary rotation diagrams of absorbing components. $N_l$ are derived from Gaussian fitting. Solid lines represent fitting results of equation~\ref{eq:bd} and dotted lines are of equation~\ref{eq:bd2}, and the derived \tex\ and $N_\textrm{tot}$ are listed in Table~\ref{table:a1a2}. Each panel presents data from all different molecular species at a specific velocity component, and the color head on each figure follow the colors of distinct components in Figure~\ref{fig:avg-fit}.}\label{fig:all-bd-app2}}
\end{figure*}

\section{Supplementary Tables}

\begin{deluxetable*}{l rrrrrrrrcl}
\tablecolumns{11}
\tabletypesize{\scriptsize}
\tablecaption{MIR1:  \thco\ Line Parameters of $v = -43$ \kms \label{tab:appex1}}
\tablehead{ \colhead{Species} & tr.s. & \multicolumn{1}{c}{\bfseries $\lambda$} & \el& \gl & \aij & \multicolumn{1}{c}{\bfseries \vLSR} &\multicolumn{1}{c}{\bfseries $\Delta v$}  & \multicolumn{1}{c}{\bfseries  \tauu$_0$}  & \multicolumn{1}{c}{\bfseries \nl} & \multicolumn{1}{c}{\bfseries $W_v$}  \\
  & & ~(\um) & ~(K)  & & ~(s\mo) & ~(\kms) & ~(\kms) & & ~($\times~10^{14}$~\cmms) & \multicolumn{1}{c}{(\kms)} }
\startdata
    \thco~$\nu$~=~0--1 &	R0	&	4.7626	&	0	&	2	&	10.9	&	-43.4	$\pm$	0.5	&	2.5	$\pm$	0.9	&	0.593	$\pm$	0.009	&	16.1	$\pm$	11	&	2.271	$\pm$	1.547\\
&	P1	&	4.7792	&	5.3	&	6	&	32.4	&	-43	$\pm$	1.4	&	2.5	$\pm$	2.3	&	0.658	$\pm$	0.005	&	53.4	$\pm$	79.9	&	2.507	$\pm$	3.752\\
&	P2	&	4.7877	&	15.9	&	10	&	21.5	&	-43.4	$\pm$	0.4	&	2.5	$\pm$	0.7	&	0.942	$\pm$	0.007	&	59.3	$\pm$	26.6	&	3.226	$\pm$	1.496\\
&	P3	&	4.7963	&	31.7	&	14	&	19.2	&	-43.5	$\pm$	0.9	&	2.5	$\pm$	1.2	&	1.147	$\pm$	0.015	&	56.8	$\pm$	59.9	&	3.432	$\pm$	3.616\\
&	R3	&	4.7383	&	31.7	&	14	&	14.8	&	-43.2	$\pm$	0.5	&	2.5	$\pm$	1	&	1.387	$\pm$	0.009	&	52.6	$\pm$	32.8	&	3.877	$\pm$	2.643\\
&	P4	&	4.805	&	52.9	&	18	&	18.2	&	-43.2	$\pm$	0.5	&	2.5	$\pm$	1	&	1.167	$\pm$	0.014	&	64	$\pm$	37.6	&	3.667	$\pm$	2.355\\
&	P7	&	4.8317	&	148	&	30	&	16.9	&	-43.7	$\pm$	0.5	&	2.4	$\pm$	0.9	&	1.096	$\pm$	0.01	&	46.9	$\pm$	35	&	3.08	$\pm$	2.299\\
&	P8	&	4.8408	&	190.3	&	34	&	16.6	&	-43.7	$\pm$	1.9	&	1.8	$\pm$	2.1	&	0.973	$\pm$	0.021	&	43.5	$\pm$	65	&	2.116	$\pm$	4.303\\
&	P9	&	4.8501	&	237.9	&	38	&	16.4	&	-43.7	$\pm$	1	&	2	$\pm$	1.3	&	0.815	$\pm$	0.009	&	39.4	$\pm$	40.1	&	2.154	$\pm$	2.670\\
&	P10	&	4.8594	&	290.7	&	42	&	16.2	&	-44.4	$\pm$	3.5	&	1.5	$\pm$	3.6	&	0.565	$\pm$	0.099	&	20.3	$\pm$	50.1	&	1.092	$\pm$	3.355\\
&	R10	&	4.6853	&	290.7	&	42	&	16.5	&	-43.7	$\pm$	0.6	&	2.2	$\pm$	0.9	&	0.728	$\pm$	0.006	&	32.9	$\pm$	24.5	&	2.259	$\pm$	1.815\\
&	P11	&	4.8689	&	348.9	&	46	&	16.1	&	-43.8	$\pm$	1	&	2.1	$\pm$	1.4	&	0.535	$\pm$	0.005	&	26.4	$\pm$	32.9	&	1.712	$\pm$	2.215\\
&	P12	&	4.8784	&	412.3	&	50	&	15.9	&	-43.4	$\pm$	1.2	&	2.2	$\pm$	1.8	&	0.434	$\pm$	0.005	&	24.1	$\pm$	33.8	&	1.606	$\pm$	2.284\\
&	R12	&	4.6711	&	412.3	&	50	&	16.8	&	-43.2	$\pm$	0.7	&	2.5	$\pm$	0.7	&	0.506	$\pm$	0.006	&	32.3	$\pm$	14.8	&	2.168	$\pm$	1.087\\
&	P13	&	4.8881	&	480.9	&	54	&	15.8	&	-43.2	$\pm$	0.6	&	2.5	$\pm$	1	&	0.353	$\pm$	0.006	&	23.3	$\pm$	15.6	&	1.569	$\pm$	1.057\\
&	R13	&	4.6641	&	480.9	&	54	&	16.9	&	-43.3	$\pm$	1.2	&	2.4	$\pm$	1.8	&	0.386	$\pm$	0.004	&	22.3	$\pm$	27.5	&	1.599	$\pm$	2.013\\
\enddata 
\tablecomments{Table~\ref{tab:appex1} is published in its entirety in the machine readable format, which lists the line parameters of all decomposed absorption line components of MIR1/MIR2 at different velocities. A portion is shown here for guidance regarding its form and content.}
\end{deluxetable*}


\begin{thebibliography}{}

\bibitem[Ag{\'u}ndez et al.(2008)]{agundez08} Ag{\'u}ndez, M., Cernicharo, J., \& Goicoechea, J.~R.\ 2008, \aap, 483, 831. doi:10.1051/0004-6361:20077927

\bibitem[Astropy Collaboration et al.(2013)]{astropy13} Astropy Collaboration, Robitaille, T.~P., Tollerud, E.~J., et al.\ 2013, \aap, 558, A33. doi:10.1051/0004-6361/201322068

\bibitem[Astropy Collaboration et al.(2018)]{astropy18} Astropy Collaboration, Price-Whelan, A.~M., Sip{\H{o}}cz, B.~M., et al.\ 2018, \aj, 156, 123. doi:10.3847/1538-3881/aabc4f

\bibitem[Bally \& Zinnecker(2005)]{bally05} Bally, J. \& Zinnecker, H.\ 2005, \aj, 129, 2281. doi:10.1086/429098

\bibitem[Barr et al.(2020)]{barr20} Barr, A.~G., Boogert, A., DeWitt, C.~N., et al.\ 2020, \apj, 900, 104. doi:10.3847/1538-4357/abab05

\bibitem[Bast et al.(2013)]{bast13} Bast, J.~E., Lahuis, F., van Dishoeck, E.~F., et al.\ 2013, \aap, 551, A118. doi:10.1051/0004-6361/201219908

\bibitem[Beltr{\'a}n \& de Wit(2016)]{beltran16} Beltr{\'a}n, M.~T. \& de Wit, W.~J.\ 2016, \aapr, 24, 6. doi:10.1007/s00159-015-0089-z

\bibitem[Bernstein et al.(1995)]{bs95} Bernstein, M.~P., Sandford, S.~A., Allamandola, L.~J., et al.\ 1995, \apj, 454, 327

\bibitem[Bernstein et al.(1997)]{bs97} Bernstein, M.~P., Allamandola, L.~J., \& Sandford, S.~A.\ 1997, Advances in Space Research, 19, 991

\bibitem[Beuther et al.(2007)]{beither07} Beuther, H., Churchwell, E.~B., McKee, C.~F., et al.\ 2007, Protostars and Planets V, 165

\bibitem[Bohlin et al.(1978)]{bohlin78} Bohlin, R.~C., Savage, B.~D., \& Drake, J.~F.\ 1978, \apj, 224, 132. doi:10.1086/156357

\bibitem[Bonnell et al.(1998)]{bonnell98} Bonnell, I.~A., Bate, M.~R., \& Zinnecker, H.\ 1998, \mnras, 298, 93. doi:10.1046/j.1365-8711.1998.01590.x

\bibitem[Bonnell et al.(2004)]{bonnell04} Bonnell, I.~A., Vine, S.~G., \& Bate, M.~R.\ 2004, \mnras, 349, 735. doi:10.1111/j.1365-2966.2004.07543.x

\bibitem[Bonnell \& Bate(2006)]{bonnell06} Bonnell, I.~A. \& Bate, M.~R.\ 2006, \mnras, 370, 488. doi:10.1111/j.1365-2966.2006.10495.x

\bibitem[Campbell et al.(1995)]{campbell95} Campbell, M.~F., Butner, H.~M., Harvey, P.~M., et al.\ 1995, \apj, 454, 831. doi:10.1086/176536

\bibitem[Cardelli et al.(1996)]{cardelli96} Cardelli, J.~A., Meyer, D.~M., Jura, M., et al.\ 1996, \apj, 467, 334. doi:10.1086/177608

\bibitem[Cesaroni et al.(2007)]{cesaroni07} Cesaroni, R., Galli, D., Lodato, G., et al.\ 2007, Protostars and Planets V, 197

\bibitem[Cushing et al.(2004)]{cushing04} Cushing, M.~C., Vacca, W.~D., \& Rayner, J.~T.\ 2004, \pasp, 116, 362

\bibitem[de Wit et al.(2011)]{dewit11} de Wit, W.~J., Hoare, M.~G., Oudmaijer, R.~D., et al.\ 2011, \aap, 526, L5. doi:10.1051/0004-6361/201016062

\bibitem[Draine \& McKee(1993)]{draine93} Draine, B.~T. \& McKee, C.~F.\ 1993, \araa, 31, 373. doi:10.1146/annurev.aa.31.090193.002105

\bibitem[Draine(2011)]{draine10} Draine, B.~T.\ 2011, Physics of the Interstellar and Intergalactic Medium by Bruce T. Draine. Princeton University Press, 2011. ISBN: 978-0-691-12214-4

\bibitem[Frost et al.(2021a)]{frost21} Frost, A.~J., Oudmaijer, R.~D., Lumsden, S.~L., et al.\ 2021, \apj, 920, 48. doi:10.3847/1538-4357/ac1741

\bibitem[Frost et al.(2021b)]{frost21b} Frost, A.~J., Oudmaijer, R.~D., de Wit, W.~J., et al.\ 2021, \aap, 648, A62. doi:10.1051/0004-6361/202039748

\bibitem[Gibb et al.(2004)]{gibb04} Gibb, E.~L., Whittet, D.~C.~B., Boogert, A.~C.~A., et al.\ 2004, \apjs, 151, 35. doi:10.1086/381182

\bibitem[Gwinn(1994)]{gwinn94} Gwinn, C.~R.\ 1994, \apj, 429, 241. doi:10.1086/174315

\bibitem[Harris et al.(2020)]{harris20} Harris, C.~R., Millman, K.~J., van der Walt, S.~J., et al.\ 2020, \nat, 585, 357. doi:10.1038/s41586-020-2649-2

\bibitem[Herbst \& van Dishoeck(2009)]{herbst09} Herbst, E. \& van Dishoeck, E.~F.\ 2009, \araa, 47, 427. doi:10.1146/annurev-astro-082708-101654

\bibitem[Hollenbach et al.(2013)]{hem13} Hollenbach, D., Elitzur, M., \& McKee, C.~F.\ 2013, \apj, 773, 70. doi:10.1088/0004-637X/773/1/70

\bibitem[Hsieh et al.(2021)]{hsieh21} Hsieh, T.-H., Takami, M., Connelley, M.~S., et al.\ 2021, \apj, 912, 108. doi:10.3847/1538-4357/abee88

\bibitem[Imai et al.(2000)]{imai00} Imai, H., Kameya, O., Sasao, T., et al.\ 2000, \apj, 538, 751. doi:10.1086/309165

\bibitem[Imai et al.(2002)]{imai02} Imai, H., Deguchi, S., \& Sasao, T.\ 2002, \apj, 567, 971. doi:10.1086/338582

\bibitem[Indebetouw et al.(2005)]{inte05} Indebetouw, R., Mathis, J.~S., Babler, B.~L., et al.\ 2005, \apj, 619, 931. doi:10.1086/426679

\bibitem[Jacob et al.(2019)]{jacob19} Jacob, A.~M., Menten, K.~M., Wiesemeyer, H., et al.\ 2019, \aap, 632, A60. doi:10.1051/0004-6361/201936037

\bibitem[Jacob et al.(2022)]{jacob22} Jacob, A.~M., Neufeld, D.~A., Schilke, P., et al.\ 2022, \apj, 930, 141. doi:10.3847/1538-4357/ac5409

\bibitem[Kauffmann et al.(2008)]{kauffmann08} Kauffmann, J., Bertoldi, F., Bourke, T.~L., et al.\ 2008, \aap, 487, 993. doi:10.1051/0004-6361:200809481

\bibitem[Kaufman \& Neufeld(1996)]{kaufman96} Kaufman, M.~J. \& Neufeld, D.~A.\ 1996, \apj, 456, 250. doi:10.1086/176645

\bibitem[Keady et al.(1988)]{keady88} Keady, J.~J., Hall, D.~N.~B., \& Ridgway, S.~T.\ 1988, \apj, 326, 832. doi:10.1086/166141

\bibitem[Kochanov et al.(2016)]{kochanov16} Kochanov, R.~V., Gordon, I.~E., Rothman, L.~S., et al.\ 2016, \jqsrt, 177, 15. doi:10.1016/j.jqsrt.2016.03.005

\bibitem[Krumholz et al.(2005)]{krum05} Krumholz, M.~R., McKee, C.~F., \& Klein, R.~I.\ 2005, \apjl, 618, L33. doi:10.1086/427555

\bibitem[Krumholz et al.(2009)]{krum09} Krumholz, M.~R., Klein, R.~I., McKee, C.~F., et al.\ 2009, Science, 323, 754. doi:10.1126/science.1165857

\bibitem[Lacy(2013)]{lacy13} Lacy, J.~H.\ 2013, \apj, 765, 130. doi:10.1088/0004-637X/765/2/130

\bibitem[Lahuis \& van Dishoeck(2000)]{lahuis00} Lahuis, F. \& van Dishoeck, E.~F.\ 2000, \aap, 355, 699

\bibitem[Loison et al.(2014)]{loison14} Loison, J.-C., Wakelam, V., Hickson, K.~M., et al.\ 2014, \mnras, 437, 930. doi:10.1093/mnras/stt1956

\bibitem[Maud et al.(2019)]{maud19} Maud, L.~T., Cesaroni, R., Kumar, M.~S.~N., et al.\ 2019, \aap, 627, L6. doi:10.1051/0004-6361/201935633

\bibitem[McKee \& Tan(2003)]{mt03} McKee, C.~F. \& Tan, J.~C.\ 2003, \apj, 585, 850. doi:10.1086/346149

\bibitem[McKee \& Ostriker(2007)]{mc07} McKee, C.~F. \& Ostriker, E.~C.\ 2007, \araa, 45, 565. doi:10.1146/annurev.astro.45.051806.110602

\bibitem[Megeath et al.(1996)]{megeath96} Megeath, S.~T., Herter, T., Beichman, C., et al.\ 1996, \aap, 307, 775

\bibitem[Megeath et al.(2005)]{megeath05} Megeath, S.~T., Wilson, T.~L., \& Corbin, M.~R.\ 2005, \apjl, 622, L141. doi:10.1086/429720

\bibitem[Menten et al.(1990)]{menten90} Menten, K.~M., Melnick, G.~J., \& Phillips, T.~G.\ 1990, Liege International Astrophysical Colloquia, 29, 243

\bibitem[Mihalas(1978)]{mihalas78} Mihalas, D.\ 1978, San Francisco: W.H. Freeman, 1978

\bibitem[Milam et al.(2005)]{milam05} Milam, S.~N., Savage, C., Brewster, M.~A., et al.\ 2005, \apj, 634, 1126. doi:10.1086/497123

\bibitem[Mitchell et al.(1990)]{mitchell90} Mitchell, G.~F., Maillard, J.-P., Allen, M., et al.\ 1990, \apj, 363, 554. doi:10.1086/169365

\bibitem[Mitchell et al.(1991)]{mitchell91} Mitchell, G.~F., Maillard, J.-P., \& Hasegawa, T.~I.\ 1991, \apj, 371, 342. doi:10.1086/169896

\bibitem[Monnier et al.(2009)]{monnier09} Monnier, J.~D., Tuthill, P.~G., Ireland, M., et al.\ 2009, \apj, 700, 491. doi:10.1088/0004-637X/700/1/491

\bibitem[Navarete et al.(2019)]{navarete19} Navarete, F., Galli, P.~A.~B., \& Damineli, A.\ 2019, \mnras, 487, 2771. doi:10.1093/mnras/stz1442

\bibitem[Neufeld \& Hollenbach(1994)]{neufeld94} Neufeld, D.~A. \& Hollenbach, D.~J.\ 1994, \apj, 428, 170. doi:10.1086/174230

\bibitem[Olguin et al.(2020)]{olguin20} Olguin, F.~A., Hoare, M.~G., Johnston, K.~G., et al.\ 2020, \mnras, 498, 4721. doi:10.1093/mnras/staa2406

\bibitem[Ormel et al.(2011)]{ormel11} Ormel, C.~W., Min, M., Tielens, A.~G.~G.~M., et al.\ 2011, \aap, 532, A43. doi:10.1051/0004-6361/201117058

\bibitem[Pontoppidan et al.(2003)]{pon03} Pontoppidan, K.~M., Fraser, H.~J., Dartois, E., et al.\ 2003, \aap, 408, 981. doi:10.1051/0004-6361:20031030

\bibitem[Purser et al.(2021)]{purser21} Purser, S.~J.~D., Lumsden, S.~L., Hoare, M.~G., et al.\ 2021, \mnras. doi:10.1093/mnras/stab747

\bibitem[Rayner et al.(2003)]{rayner03} Rayner, J.~T., Toomey, D.~W., Onaka, P.~M., et al.\ 2003, \pasp, 115, 362. doi:10.1086/367745


\bibitem[Rayner et al.(2022)]{rayner22} Rayner, J., Tokunaga, A., Jaffe, D., et al.\ 2022, \pasp, 134, 015002. doi:10.1088/1538-3873/ac3cb4


\bibitem[Roche \& Aitken(1984)]{roche84} Roche, P.~F. \& Aitken, D.~K.\ 1984, \mnras, 208, 481. doi:10.1093/mnras/208.3.481

\bibitem[Rodgers \& Williams(1974)]{rodgers74} Rodgers, C.~D. \& Williams, A.~P.\ 1974, \jqsrt, 14, 319. doi:10.1016/0022-4073(74)90113-7

\bibitem[Rod{\'o}n et al.(2008)]{rodon08} Rod{\'o}n, J.~A., Beuther, H., Megeath, S.~T., et al.\ 2008, \aap, 490, 213. doi:10.1051/0004-6361:200810158

\bibitem[Rosen \& Krumholz(2020)]{rosen20} Rosen, A.~L. \& Krumholz, M.~R.\ 2020, \aj, 160, 78. doi:10.3847/1538-3881/ab9abf

\bibitem[Sheffer et al.(2008)]{sheffer08} Sheffer, Y., Rogers, M., Federman, S.~R., et al.\ 2008, \apj, 687, 1075. doi:10.1086/591484

\bibitem[Sofia et al.(1997)]{sofia97} Sofia, U.~J., Cardelli, J.~A., Guerin, K.~P., et al.\ 1997, \apjl, 482, L105. doi:10.1086/310681


\bibitem[Suri et al.(2021)]{suri21} Suri, S., Beuther, H., Gieser, C., et al.\ 2021, \aap, 655, A84. doi:10.1051/0004-6361/202140963

\bibitem[Tielens \& Hollenbach(1985)]{tielens85} Tielens, A.~G.~G.~M. \& Hollenbach, D.\ 1985, \apj, 291, 722. doi:10.1086/163111

\bibitem[Tielens(2005)]{tielens05} Tielens, A.~G.~G.~M.\ 2005, The Physics and Chemistry of the Interstellar Medium, by A. G. G. M. Tielens, pp. . ISBN 0521826349. Cambridge, UK: Cambridge University Press,  2005.

\bibitem[Tielens(2021)]{tielens21} Tielens, A.~G.~G.~M.\ 2021, Molecular Astrophysics. Cambridge: Cambridge University Press. doi:10.1017/9781316718490

\bibitem[van der Tak et al.(2000)]{vdt00} van der Tak, F.~F.~S., van Dishoeck, E.~F., Evans, N.~J., et al.\ 2000, \apj, 537, 283. doi:10.1086/309011

\bibitem[van der Tak et al.(2003)]{vdt03} van der Tak, F.~F.~S., Boonman, A.~M.~S., Braakman, R., et al.\ 2003, \aap, 412, 133. doi:10.1051/0004-6361:20031409

\bibitem[van der Tak et al.(2005)]{vdt05} van der Tak, F.~F.~S., Tuthill, P.~G., \& Danchi, W.~C.\ 2005, \aap, 431, 993. doi:10.1051/0004-6361:20041595

\bibitem[Vacca et al.(2003)]{vacca03} Vacca, W.~D., Cushing, M.~C., \& Rayner, J.~T.\ 2003, \pasp, 115, 389. doi:10.1086/346193

\bibitem[Villanueva et al.(2018)]{vill18} Villanueva, G.~L., Smith, M.~D., Protopapa, S., et al.\ 2018, \jqsrt, 217, 86. doi:10.1016/j.jqsrt.2018.05.023

\bibitem[Vinogradoff et al.(2013)]{vino13} Vinogradoff, V., Fray, N., Duvernay, F., et al.\ 2013, \aap, 551, A128

\bibitem[Virtanen et al.(2020)]{virtanen20} Virtanen, P., Gommers, R., Oliphant, T.~E., et al.\ 2020, Nature Methods, 17, 261. doi:10.1038/s41592-019-0686-2

\bibitem[Wang et al.(2012)]{wang12} Wang, Y., Beuther, H., Zhang, Q., et al.\ 2012, \apj, 754, 87. doi:10.1088/0004-637X/754/2/87

\bibitem[Wang et al.(2013)]{wang13} Wang, K.-S., Bourke, T.~L., Hogerheijde, M.~R., et al.\ 2013, \aap, 558, A69. doi:10.1051/0004-6361/201322087

\bibitem[Wouterloot et al.(2008)]{wouterloot08} Wouterloot, J.~G.~A., Henkel, C., Brand, J., et al.\ 2008, \aap, 487, 237. doi:10.1051/0004-6361:20078156

\bibitem[Wilson \& Rood(1994)]{wilson94} Wilson, T.~L. \& Rood, R.\ 1994, \araa, 32, 191. doi:10.1146/annurev.aa.32.090194.001203

\bibitem[Wilson et al.(2003)]{wilson03} Wilson, T.~L., Boboltz, D.~A., Gaume, R.~A., et al.\ 2003, \apj, 597, 434. doi:10.1086/378233

\bibitem[Wolfire \& Cassinelli(1987)]{wolfire87} Wolfire, M.~G. \& Cassinelli, J.~P.\ 1987, \apj, 319, 850. doi:10.1086/165503

\bibitem[Zinnecker \& Yorke(2007)]{zin07} Zinnecker, H. \& Yorke, H.~W.\ 2007, \araa, 45, 481. doi:10.1146/annurev.astro.44.051905.092549

\end{thebibliography}
\end{document}